\documentclass[%
 reprint,
superscriptaddress,
preprintnumbers,
nofootinbib,
 amsmath,amssymb,
 aps,
]{revtex4-2}

\usepackage{graphicx}% Include figure files
\usepackage{dcolumn}% Align table columns on decimal point
\usepackage{bm}% bold math
\usepackage{hyperref}% add hypertext capabilities
\hypersetup{
    unicode=false,          % non-Latin characters 
    pdftoolbar=true,        % show Acrobat
    pdfmenubar=true,        % show Acrobat 
    pdffitwindow=false,     % window fit to page when opened
    pdfstartview={FitH},    % fits the width of the page to the window
    pdftitle={Data-driven classification of topological lasing modes},    % title
    pdfauthor={Wong, Olthaus, Bracht, Reiter, Oh},     % author
    pdfsubject={},   % subject of the document
    pdfcreator={},   % creator of the document
    pdfproducer={}, % producer of the document
    pdfkeywords={} {} {}, % list of keywords
    pdfnewwindow=true,      % links in new window
    colorlinks=true,       % false: boxed links; true: colored links
    linkcolor=blue, %red,          % color of internal links (change box color with linkbordercolor)
    citecolor=blue,        % color of links to bibliography
    filecolor=magenta,      % color of file links
    urlcolor=blue,           % color of external links
	breaklinks=true
} 

\usepackage{amssymb}
\usepackage{amsmath}
\usepackage{amsfonts}
\usepackage{xspace}
\usepackage{bm,bbm}
\usepackage{relsize}
\usepackage{siunitx}
\usepackage{xcolor}

\newcommand{\norm}[1]{\lVert#1\rVert}

\begin{document}
\title{A machine learning approach to drawing phase diagrams of topological lasing modes}

\author{Stephan Wong}
\affiliation{School of Physics and Astronomy, Cardiff University, Cardiff CF24 3AA, UK}

\author{Jan Olthaus}
 \affiliation{Institut f\"ur Festk\"orpertheorie, Universit\"at M\"unster, 48149 M\"unster, Germany}
 
\author{Thomas K. Bracht}
\affiliation{Institut f\"ur Festk\"orpertheorie, Universit\"at M\"unster, 48149 M\"unster, Germany}
 
\author{Doris E. Reiter}
\affiliation{Institut f\"ur Festk\"orpertheorie, Universit\"at M\"unster, 48149 M\"unster, Germany}
\affiliation{Condensed Matter Theory, Department of Physics, TU Dortmund, 44221 Dortmund, Germany}

\author{Sang Soon Oh}
\email[Email: ]{ohs2@cardiff.ac.uk}
\affiliation{School of Physics and Astronomy, Cardiff University, Cardiff CF24 3AA, UK}

\date{\today}

\begin{abstract}
Identifying phases and analyzing the stability of dynamic states are ubiquitous and important problems which appear in various physical systems.
Nonetheless, drawing a phase diagram in high-dimensional and large parameter spaces has remained challenging.  
Here, we propose a data-driven method to derive the phase diagram of lasing modes in topological insulator lasers. 
The classification is based on the temporal behaviour of the topological modes obtained via numerical integration of the rate equation. 
A semi-supervised learning method is used and an adaptive library is constructed in order to distinguish the different topological modes present in the generated parameter space. 
The proposed method successfully distinguishes the different topological phases in the Su-Schrieffer-Heeger (SSH) lattice with saturable gain.
This demonstrates the possibility of classifying the topological phases without needing for expert knowledge of the system and may give valuable insight into the fundamental physics of topological insulator lasers via reverse engineering.
\end{abstract}

% \pacs{03.65.Vf, 03.65.Fd, 42.55.Tv, 78.67.Pt, 73.21.-b,}
% \keywords{}

\maketitle

%=========================
%=========================
%=========================
%=========================

\section{Introduction}

Over the last few years, significant research efforts have been made on photonic topological insulator (PTI) lasers.
While the efforts have been concentrated on the spatial stability of the topologically protected edge modes, namely on the existence of such topological edge modes in non-Hermitian PTIs~\cite{Schomerus2013, Poli2015, Bahari2017, Leykam2017, Zhang2018, Parto2018, Takata2018, Kawabata2019, Shao2020}, the temporal stability has not been the focus of interest so far~\cite{Longhi2018, Longhi2018a, Malzard2018, Malzard2018, Malzard2018a, Wong2021, Gong2020}.
Due to the non-linear nature of PTI lasers, the temporal stability is an important characteristic to take into account for experimental demonstrations and real-life applications. 
Although the spatial stability of the topological modes, \emph{i.e.}, its robustness, may be guaranteed in active non-Hermitian PTIs, unstable behaviour may be present in the time domain.
In this regard, the temporal dynamics of the topologically protected modes have been studied~\cite{Longhi2018, Malzard2018, Malzard2018a}, mainly using linear stability analysis.
It is, however, a challenging task to apply the same approach to more complex structures because of the high-dimensional phase space and parameter space as well as the lack of analytical solutions~\cite{Longhi2018a}.

Machine learning (ML) can be advantageous for the theoretical study of the stability of PTIs, which requires repetitive numerical simulations for several varying parameters.
ML is a data-based method which can be implemented with different strategies, and the most appropriate ML strategy depends on the dataset under study. 
For instance, a supervised learning strategy relies on labelled data, a dataset of input-output pairs.
This has been utilized in topological photonics to draw topological phase diagrams~\cite{Araki2019}, calculate topological invariants~\cite{Zhang2018}, or explore topological band structures~\cite{Peano2021}. 
On the other hand, an unsupervised learning strategy consists of extracting information from the dataset for which we do not have labels.
This is used for dimensional reductions by keeping only the main features of the high-dimensional structure of the dataset or for clustering problems from which the data is divided into different types~\cite{FabianPedregosa2011}.
For instance, this has been successful in obtaining the phase transition in the Ising model~\cite{Wang2016a}, and clustering Hamiltonians that belong to the same symmetry classes~\cite{Scheurer2020}.

In the unsupervised learning strategy, modal decomposition is a common and successful method which reduces the analysis of very high-dimensional data to a set of relatively few modes.
Among the modal decomposition methods, principal component analysis (PCA) is a method which derives the eigenmodes or the main features based on their variance in the data~\cite{Jolliffe2005}.
These eigenmodes can then be utilized as a basis to represent the dataset~\cite{Wright2009}.
This reduced-order model method has been extended to identify distinct non-linear regimes~\cite{Brunton2014, Proctor2014, Fu2014, Brunton2016a, Bright2016, Kramer2017, Ozan2021} by constructing a library composed of representatives of these regimes: this is known as representation classification.
Nevertheless, the preliminary identification of the regimes composing the library and the construction of the library is a manual process and requires expert knowledge of the complex system.

In this paper, we propose a representation classification method to study the spatio-temporal dynamics of non-linear topological systems.
The results will be based on the phase diagram of the Su-Schrieffer-Heeger (SSH) lattice~\cite{Su1979} with a domain wall and with saturable gain~\cite{Malzard2018, Malzard2018a}.
To remove the necessity of the required expertise on the complex system, we present an algorithm which constructs an appropriate library of the different phases automatically.
For this goal, we propose two approaches: a top-down approach in which the library has numerous phases that are merged into the equivalent phases, and a bottom-up approach in which the library is completed on the fly to get the most accurate classification.
Via reverse engineering, our proposed method can be used as a tool to find novel topological lasing modes in more complicated settings.
For given rate equations of a lasing system, one would only need to integrate the differential equations in the desired parameter space region and then apply the adaptive representation classification to obtain the phase diagram.

%=========================
%=========================
%=========================
%=========================

\section{Results}

%=========================
%=========================
%=========================

\subsection{Phase diagram of the SSH Model}

\begin{figure}[!]
\center
\includegraphics[width=\columnwidth]{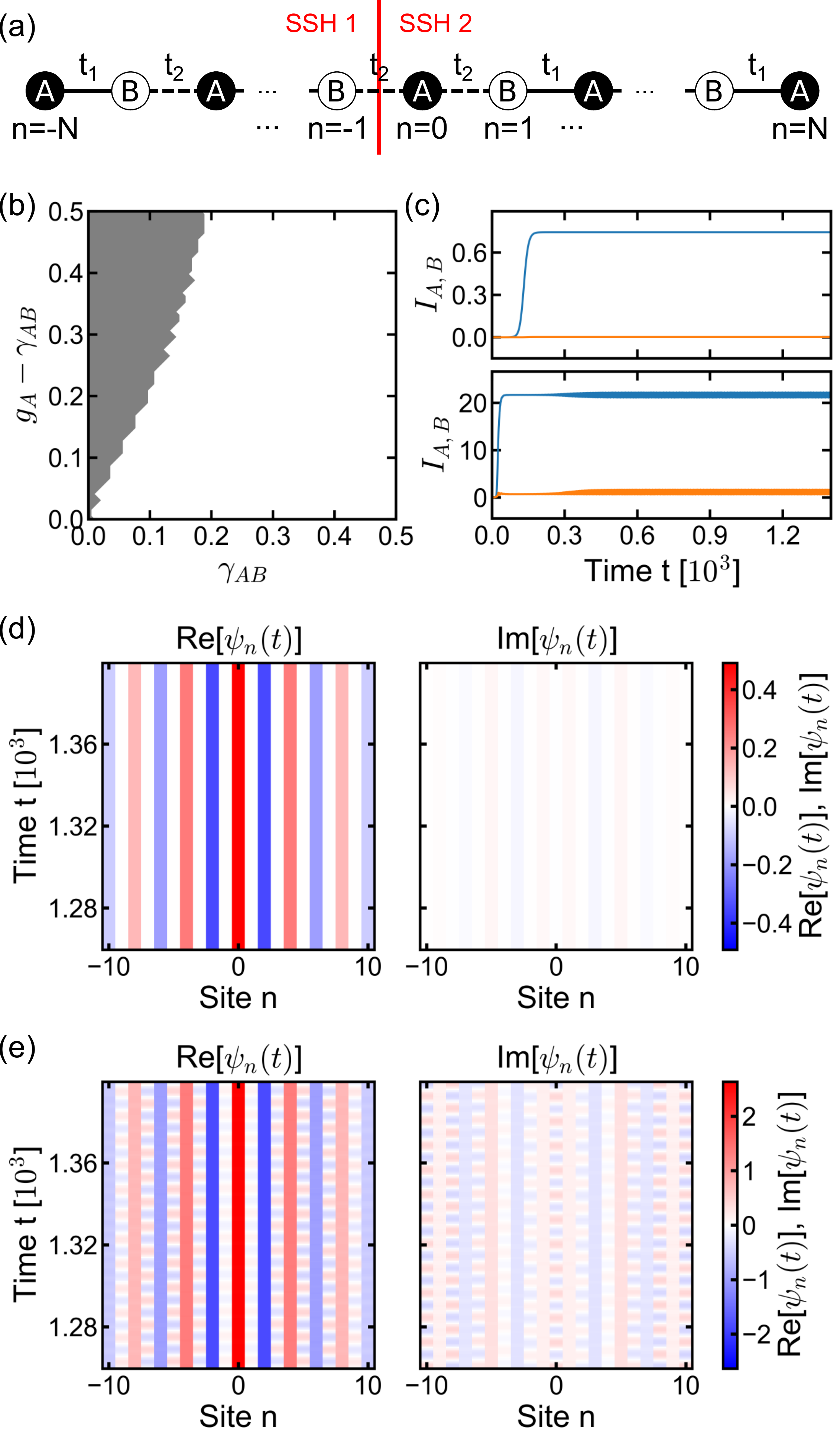}
\caption{
\textbf{Phase diagram of the domain-wall-type SSH lattice with saturable gain.}
(a) Schematic of the domain-wall-type SSH lattice considered. 
The vertical red line is a guide to the eye for the domain wall between the two SSH lattices, SSH 1 and SSH 2.
(b) Phase diagram of the domain-wall-type SSH lattice with saturable gain and linear loss on the A and B sublattices.
The white and grey areas correspond to the non-oscillating and oscillating topological phases, respectively.
(c) Representative total intensity $I_A$ (and $I_B$) of the A (and B) sublattice in blue (and orange) for the non-oscillating and oscillating topological lasing mode on the top and bottom panel, respectively.
Spatio-temporal dynamics of the (d) non-oscillating and (e) oscillating topological lasing modes.
%
%The lattice is made of $N_s=21$ sites ($N=10$), and the coupling and gain parameters chosen are $t_1 = 1$, $t_2 = 0.7$ and $g_B = 0$.
The non-oscillating and oscillating topological modes displayed are chosen at ($\gamma_{AB}, g_A-\gamma_{AB}) = (0.48, 0.06)$ and $(0.16, 0.44)$, respectively.
}
\label{fig:ssh_sat_gain}
\end{figure}

As a toy model, we will consider the domain-wall-type SSH lattice with saturable gain [Fig.~\ref{fig:ssh_sat_gain}(a)].
The system has a domain wall, at the A site $n=0$, which separates two SSH lattices, namely lattices composed of two sites per unit cell, A and B, and characterized by intra- and inter-unit cell couplings, $t_\text{intra}$ and $t_\text{inter}$, respectively.
$t_\text{intra}=t_1$ and $t_\text{inter}=t_2$ ($t_\text{intra}=t_2$ and $t_\text{inter}=t_1$) for the lattice on the left (right) side of the domain wall, \emph{i.e.}, the sites with $n < 0$ ($n \geq 0$).
The dynamics of the system $\psi(t) = \left( \psi_{-N}(t), \ldots, \psi_{-1}(t), \psi_0(t), \psi_1(t), \ldots, \psi_{N}(t) \right) \equiv x(t)$, with $N_s = 2 N + 1$ sites, reads, for $n = -N, \ldots, N$:
\begin{align}
\begin{split}
\label{eq:rate_equation_ssh}
i \frac{d \psi_n}{dt} & = i \left( \frac{g_n}{1+|\psi_n|^2} - \gamma_n \right) \psi_n \\[1.ex]
& \qquad + t_\text{intra,n+1} \psi_{n+1} + t_\text{inter,n-1} \psi_{n-1}
\end{split}
\end{align}
where $\psi_n$ is the amplitude of the $n$-th site.
%
%t_\text{intra,n}$ and $t_\text{inter,n}$ stand for the intra- and inter-unit cell couplings, respectively.
%
%$t_\text{intra,n}=t_1$ and $t_\text{inter,n}=t_2$ for $n < 0$, and $t_\text{intra,n}=t_2$ and $t_\text{inter,n}=t_1$ for $n \geq 0$.
%
$g_n$ and $\gamma_n$ are the linear gain and linear loss at the $n$-th site, respectively.
Using explicitly the amplitudes $a_p$ and $b_p$ of the $A$ and $B$ sites on the $p$-th unit cell, respectively, in $\psi(t) = \left( \ldots, a_p(t), b_p(t), \ldots \right)$, Eq.~\eqref{eq:rate_equation_ssh} can be re-written as:
\begin{align}
\label{eq:rate_equation_ssh_A}
i \frac{d a_p}{dt} &= i \left( \frac{g_A}{1+|a_p|^2} - \gamma_A \right) a_p + t_\text{intra} b_p + t_\text{inter} b_{p-1}
, \\
\label{eq:rate_equation_ssh_B}
i \frac{d b_p}{dt} &= i \left( \frac{g_B}{1+|b_p|^2} - \gamma_B \right) b_p + t_\text{intra} a_p + t_\text{inter} a_{p+1}
,
\end{align}
where $g_\sigma$ and $\gamma_\sigma$ are the linear gain and linear loss at the site $\sigma = A, B$.

In the passive setting ($g_A=g_B=0$, $\gamma_A=\gamma_B=0$), this configuration, with $t_1 > t_2$, is known to give a single topologically protected zero-energy (non-oscillating) mode localized at the domain wall and with non-vanishing amplitudes only on the A sites~\cite{Asboth2016}.
This is due to the bulk-boundary correspondence at the domain wall between trivial and non-trivial topological SSH lattices.
Indeed, an SSH lattice is topologically trivial (non-trivial) if the intra-unit cell coupling is lower (greater) than the inter-unit cell coupling.

In the active setting, it has been shown that the topological phase of the system depends on the gain and coupling parameters~\cite{Malzard2018, Malzard2018a}.
%
%As shown in Fig.~\ref{fig:ssh_sat_gain}(b), we can obtain the phase diagram for a lattice composed of $N_s=21$ sites ($N=10$) with $t_1 = 1$, $t_2 = 0.7$, $g_B = 0$ by varying $g_A$, and $\gamma_A = \gamma_B \equiv \gamma_{AB}$ in the parameter space.
%
%For example, using the same parameter values as in Ref.~\ref{Malzard2018a}, \emph{i.e} $t_1 = 1$, $t_2 = 0.7$ and $g_B = 0$, Figure~\ref{fig:ssh_sat_gain}(b) shows the phase diagram, by varying $g_A$ and $\gamma_A = \gamma_B \equiv \gamma_{AB}$ in the parameter space, for a lattice composed of $N_s=21$ sites ($N=10$).
%
As parameters, we use the values from Ref.~\cite{Malzard2018a} with $t_1 = 1$, $t_2 = 0.7$, $g_B = 0$ and $\gamma_A = \gamma_B \equiv \gamma_{AB}$.
Figure~\ref{fig:ssh_sat_gain}(b) shows the phase diagram, by varying $g_A$ and $\gamma_{AB}$ in the parameter space, for a lattice composed of $N_s=21$ sites ($N=10$).
In this configuration, the system has two distinct topological phases: a non-oscillating phase [white area in Fig.~\ref{fig:ssh_sat_gain}(b)] and an oscillating phase [grey area in Fig.~\ref{fig:ssh_sat_gain}(b)].
The dynamics of the two topological phases can be visualized by plotting the total intensity $I_A = \sum_p |a_p|^2$ ($I_B = \sum_p |b_p|^2$) of the A (B) sites in Fig.~\ref{fig:ssh_sat_gain}(c) as in Ref.~\cite{Malzard2018, Malzard2018a}. 
Alternatively, more details can be understood by plotting the space-time dynamics of the topological modes as shown in Figs.~\ref{fig:ssh_sat_gain}(d) and \ref{fig:ssh_sat_gain}(e).
The non-oscillating phase is similar to the zero-energy mode in the passive SSH lattice. 
We can see in Fig.~\ref{fig:ssh_sat_gain}(d) that the mode is localized at the interface and has the majority of its amplitudes on the A sublattice.
On the other hand, the system with saturable gain exhibits a new topological phase with no counterpart in the passive setting.
The new topological mode is characterized by an edge mode at the domain wall with an oscillating behaviour of the amplitudes on the A and B sites, as shown in Fig.~\ref{fig:ssh_sat_gain}(e).

The classification of the new topological phases in non-linear systems requires, so far, an expert knowledge of the given non-linear systems, for example, the known results derived in Ref.~\cite{Malzard2018}.
In fact, the phase diagram in Fig.~\ref{fig:ssh_sat_gain}(b) has been obtained solely by the fast Fourier transform of the time series in the parameter space.
Thus, the main aim of this paper is to develop a tool to explore the topological phases of PTI lasers in more complicated settings for which we have little knowledge.

The phase diagram shown in Fig.~\ref{fig:ssh_sat_gain}(b) will serve as a reference for our proposed method.
The dataset we will utilize throughout this paper is composed of about $1000$ samples which are randomly generated from the same coupling and gain parameters' range as in Fig.~\ref{fig:ssh_sat_gain}(b).
The coupled-mode equations [Eqs.~\eqref{eq:rate_equation_ssh_A} and \eqref{eq:rate_equation_ssh_B}] are integrated using the fourth-order Runge-Kutta method and with $a_p(t=0) = b_p(t=0) = 0.01$, $\forall p$, as an initial condition.
Although the integration has been performed using a fixed time step $dt = 0.01$ until a final time at $t = 1400$, only $2000$ time snapshots are uniformly retrieved in order to keep the time series at a reasonable size.
For the parameters given above, this sample rate leaves about $10$ time steps per period for the oscillating regime case [Fig.~\ref{fig:ssh_sat_gain}(c)].
The phase diagram is then obtained solely from the time series of the states within the given parameter space. 
%

%=========================
%=========================
%=========================

\subsection{Representation classification method}

To classify topological lasing modes based on their distinct non-linear regimes, we use a representation classification method~\cite{Brunton2014, Bright2016, Kramer2017}.
The general idea of representation classification relies on the assumption, and common situations, that the dynamics of a high-dimensional system evolves on a low-dimensional attractor such as fixed points or periodic orbits~\cite{Cross1993}.
The low-dimensional structure of the attractor allows for a reduced-order model that accurately approximates the underlying behaviour of the system: the dynamics of the complex system can thus be written using a basis that spans the low-dimensional space.
Representation classification consists of constructing a library of appropriate basis, representative of the dynamical regimes of interest, and only then employ a filtering strategy to identify the regime corresponding to a given unknown time series.
In the following, we will use the term ``regime" to denote the different dynamical behaviours or the different topological phases in the non-linear SSH lattice with saturable gain.
%, here the SSH lattice with saturable gain.
%
Besides, for convenience, we will plot only the total intensity on the A ($I_A = \sum_p |a_p|^2$) and B ($I_B = \sum_p |b_p|^2$) sublattices to represent the given regimes.
Nevertheless, the time series of the complex amplitudes at each site will be considered for the construction of the library.

As is common in complex dynamical systems, the dynamics of a system close to an attractor lie in a low-dimensional space.
This means that a given spatio-temporal dynamics, denoted by the vector $x(t)$, can be approximately written in terms of a basis $\Phi = \{ \phi_i \}_{i=1,\ldots,D}$ spanning the low $D$-dimensional space, namely:
\begin{equation}
\label{eq:basis_decomposition_single}
x(t) 
\approx \sum_{i=1}^D \phi_i \beta_i(t) 
= \Phi \beta(t)
\end{equation}
where $\beta_i$ are the weighted coefficients in the above linear combination of basis states $\phi_i$.
%
%$x(t)$, at a given time $t$, is a vector column of size $N_s$ with $N_s$ the size of the spatial grid, \emph{i.e.}, the number of sites in our SSH lattice for instance.
%
Using the terminology used in the literature~\cite{Brunton2014, Bright2016, Kramer2017}, $x(t)$ will, in the following, be referred to the state measured at time $t$.

However, one of the main characteristics of non-linear systems is the drastically different dynamical behaviours with respect to the system's parameters.
Therefore, the reduced-order modelling strategy using a single representative basis, \emph{i.e.}, corresponding to a single regime, is bound to fail.
Instead of finding a global basis, we here construct a set of local bases, \emph{i.e.}, construct a library composed of the bases of each non-linear regime of interest:
\begin{equation}
\mathcal{L} = \{ \Phi_1, \cdots, \Phi_J \} = \{ \phi_{j,i} \}_{j=1,\ldots,J \, , \, i=1,\ldots,D}
,
\end{equation}
where $J$ is the number of regimes, $\Phi_j$'s are the bases of each of the dynamical regime $j$, and $\phi_{j,i}$'s are the corresponding basis states.
This is the supervised learning part of the method, from which the data-driven method attempts to capture the dynamics of the system in the reduced-order model.
Therefore, the library $\mathcal{L}$ contains the representative basis of each regime of interest, and corresponds to an overcomplete basis that approximates the dynamics of the system across the given parameter space.
A better approximation of $x(t)$, instead of using Eq.~\eqref{eq:basis_decomposition_single} for a single basis regime, then reads:
\begin{equation}
x(t) 
\approx \sum_{j=1}^J \sum_{i=1}^D \phi_{j,i} \beta_{j,i}(t) 
= \sum_{j=1}^J \Phi_j \beta_j(t)
\end{equation}
where $\beta_{j,i}$ are the weighted coefficients in the above linear combination in the overcomplete basis library $\mathcal{L}$.
It is worth noting that the library modes $\phi_{j,i}$ are not orthogonal to each other, but instead orthogonal in groups of modes for each different regime $j$.

Throughout this paper, the bases used for constructing the library $\mathcal{L}$ will be generated by using a time-augmented dynamical mode decomposition (aDMD) method~\cite{Kramer2017} to consider both the spatial and temporal behaviours (see Supplementary section~SI for additional information).

Here, we use a classification scheme based on a simple hierarchical strategy~\cite{Kramer2017}.
The regime classification approach is fundamentally a subspace identification problem, where each regime is represented by a different subspace.
Given the state $x(t_i : t_{i+N_w})$ measured within the time window $[t_i,t_{i+N_w}]$, with $N_w$ the time step window size, the correct regime $j^*$ is identified as the corresponding subspace in the library $\mathcal{L}$ closest to the measurement in the $L2$-norm sense~\cite{Kramer2017}.
In other words, the classification strategy is to find the subspace that maximises the projection of the measurement onto the regime subspace:
\begin{equation}
\label{eq:class}
j^* = \arg \max_{j=1,\ldots,J} \norm{P_j x(t_i : t_{i+N_w})}_2
,
\end{equation}
where $P_j$ is a projection operator given by:
\begin{equation}
P_j = \Phi_j \Phi_j^+
\end{equation}
with $\Phi_j^+$ being the pseudo-inverse of $\Phi_j$, $\norm{\cdot}_2$ the $L2$-norm of a vector, $\norm{v}_2 := \sqrt{ \sum_i |v_i|^2 }$, and $\arg \max$ the function that returns the index of the maximum value.
%

%One should note that sparse measurement might be desirable if the data collection is too expensive, for example if $N_s$ is very large.
%
%A slight change in the methodology is then needed and more details are explained in Supplementary section~\ref{sect_supp:sparse_measurement}.
%
For the 1D system exemplified in this paper, the data collection is not too expensive because $N_s$ is reasonable.
However, if $N_s$ is very large, sparse measurement might be desirable and a slight change in the methodology is then needed as explained in Supplementary section~SII.

%=========================
%=========================
%=========================

\subsection{Phase Diagrams}

In the following we will draw phase diagrams using different approaches to the representation classification.
In the phase diagrams we will mark the different identified regimes by the color of the dots, where the (dark or light) blue dots always mark the oscillating regime, the green dots the non-oscillating regime, the red dots the transient regime and the orange dots the transition regime.
%and other colors typically correspond to transient or transition regimes.
%
The white and grey areas are overlays of the referenced phase diagram obtained in Fig.~\ref{fig:ssh_sat_gain}.
The aDMD bases have been generated with $N_w = 25$ from the time series starting at the $1800$-th time step.

%=========================
%=========================

%\subsection{Representation classification from fixed library basis}
\subsubsection{Fixed library}

\begin{figure}
\center
\includegraphics[width=\columnwidth]{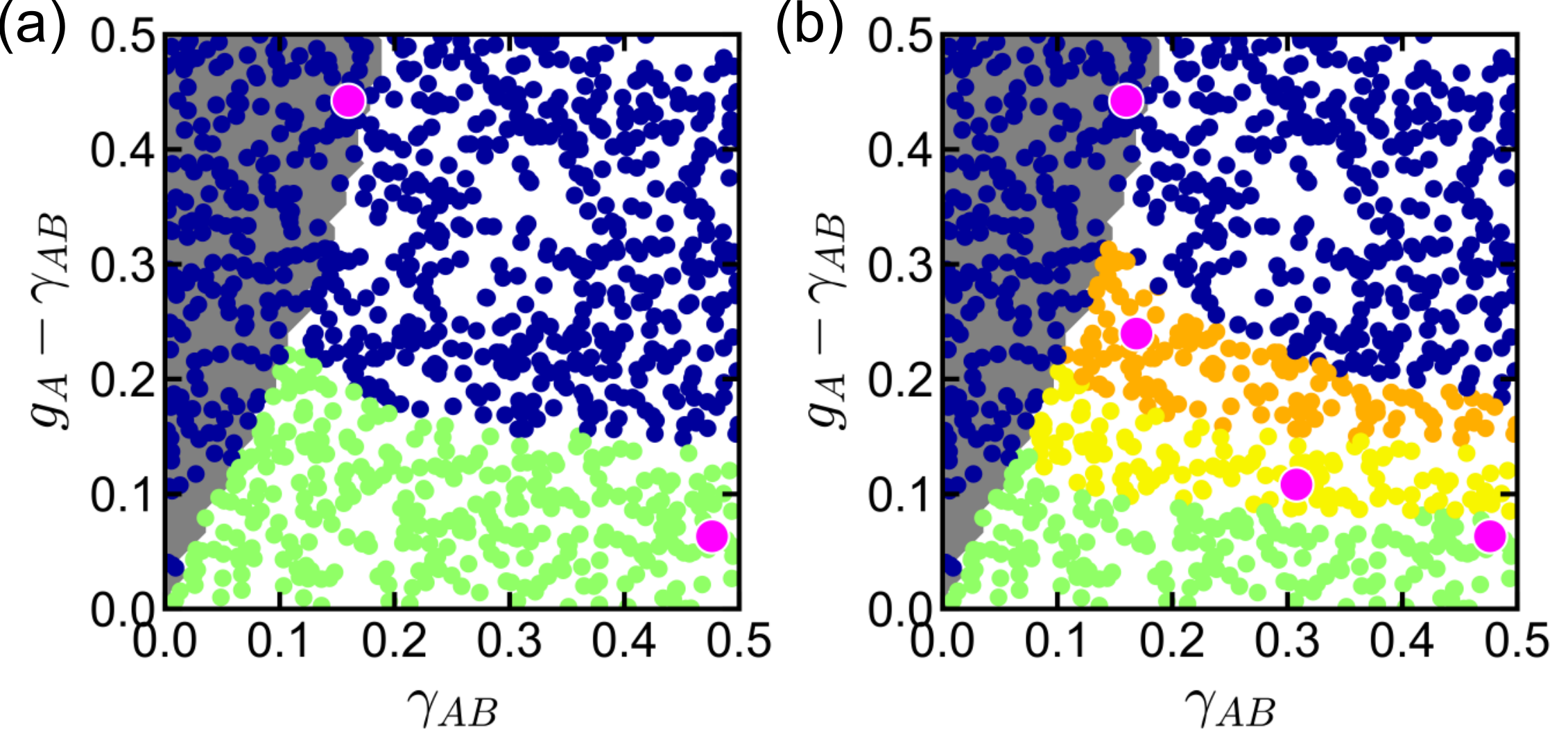}
\caption{
%\textbf{Phase diagrams derived based on representation classification.}
\textbf{Representation classification based on a fixed library.}
Phase diagram obtained from the library composed of (a) two regimes (one non-oscillating and one oscillating) and (b) four regimes (three non-oscillating and one oscillating).
The library is constructed by the magenta dots located at
($\gamma_{AB}, g_A-\gamma_{AB}) = (0.48, 0.06)$ and $(0.16, 0.44)$, 
and for (b) additionally at $(0.31, 0.11)$ and $(0.17, 0.24)$. 
}
\label{fig:phase_diagram}
\end{figure}

%
%We now draw the phase diagram of the SSH lattice with saturable gain using the representation classification described in the previous section. 
%
Figure~\ref{fig:phase_diagram}(a) displays the phase diagram derived from a library basis made of the two topological regimes known from Fig.~\ref{fig:ssh_sat_gain}.
These two topological modes, used for the construction of the library, are represented by the two magenta dots.
They are randomly chosen from the known regimes' region in Fig.~\ref{fig:ssh_sat_gain}(b) and the details of the resulted phase diagrams is dependent from that random choice.
The remaining coloured dots in the plot represent the identified regime $j^*$ [Eq.~\eqref{eq:class}] of each sample depicted by the dots in the parameter space.
%
%The grey and white areas are overlays of the referenced phase diagram [Fig.~\ref{fig:ssh_sat_gain}(b)].
%
However, we can see that the phase diagram fails to correctly predict all the dynamical behaviours.
Indeed, we observe that many time series are not correctly identified.
Using a different choice of topological regimes in the parameter space to construct the library could be a solution to find a better phase diagram, but our attempts only showed marginal improvement of the agreement.
Testing every possible choices in the dataset with no guarantee of finding the accurate phase diagram is therefore not a practical solution.
%
%Hence, testing all possible choice of topological modes to construct the library is not a practical solution.
%
Instead, using four bases in the library instead of two bases, or equivalently considering four regimes from the given parameter space, the phase diagram in Fig.~\ref{fig:phase_diagram}(b) shows better results: The identified oscillating and non-oscillating regimes are more separated and have a better fitting with the referenced phase diagram, even though they belong to a distinct regime index $j^*$.
Merging the three oscillating regimes present in the library would then give a more accurate derived phase diagram.
Therefore, Figure~\ref{fig:phase_diagram} suggests that increasing the number of bases in the library $\mathcal{L}$ and merging some of them might help to get closer to the desired phase diagram, as we will see in the later sections.
%

%=========================
%=========================

%\subsection{Representation classification from top-down adaptive library basis}
\subsubsection{Top-down adaptive library}

In the previous section, the construction of the library $\mathcal{L}$ was a manual process from which we already knew the different dynamical regimes.
This, however, requires prior knowledge of the complex system considered.
The strategy, here, is to adaptively construct the library based on the given data samples.
Here, we employ a top-down approach in which we start with too many samples for the construction of the library, and then reduce the library size by merging some of them.
Based on some measures in the decision process, this automated construction of the library thus removes the manual construction of the regimes.

The underlying assumption of the classification scheme is based on the dissimilarity between the subspace of different regimes.
We thus propose to consider regimes that are similar as equivalent regimes.
This would, for instance, help us to merge the three phases in the non-oscillating region in Fig.~\ref{fig:phase_diagram}(b), and consider them as a single regime.
In other words, the regimes $i$ and $j$ are said to be equivalent, denoted by $i \sim j$, if the fraction of information retained after the projection onto each other, $\gamma_{ij} \in [0,1]$, is high enough:
\begin{equation}
\label{eq:top_down}
\gamma_{ij} > \gamma_{th}
,
\end{equation}
where $\gamma_{th} \in [0,1]$ is the hyper-parameter which decides the threshold value for merging different regimes and $\gamma_{ij}$ is the subspace alignment given by:
\begin{equation}
\label{eq:gamma_ij}
\gamma_{ij} := \frac{\norm{P_i P_j}_F ^2}{\norm{P_i}_F \norm{P_j}_F}
,
\end{equation}
with $\norm{\cdot}_F$ the Frobenius norm of a matrix, $\norm{M}_F := \sqrt{ \sum_{ij} |M_{ij}|^2 }$.
Importantly, the relation Eq.~\eqref{eq:top_down} is numerically computed in such a way that the transitivity property of the equivalence relation is satisfied, namely that if $i \sim j$ and $j \sim k$ then $i \sim k$.
The relation Eq.~\eqref{eq:top_down} is then indeed an equivalence relation because the reflexive ($i \sim i$) and symmetric ($i \sim j \Rightarrow j \sim i$) property of the relation is automatically satisfied from the definition of $\gamma_{ij}$ [Eq.~\eqref{eq:gamma_ij}].
Supplementary section~SIII gives more details on the top-down library generation principle.

\begin{figure}
\includegraphics[width=\columnwidth]{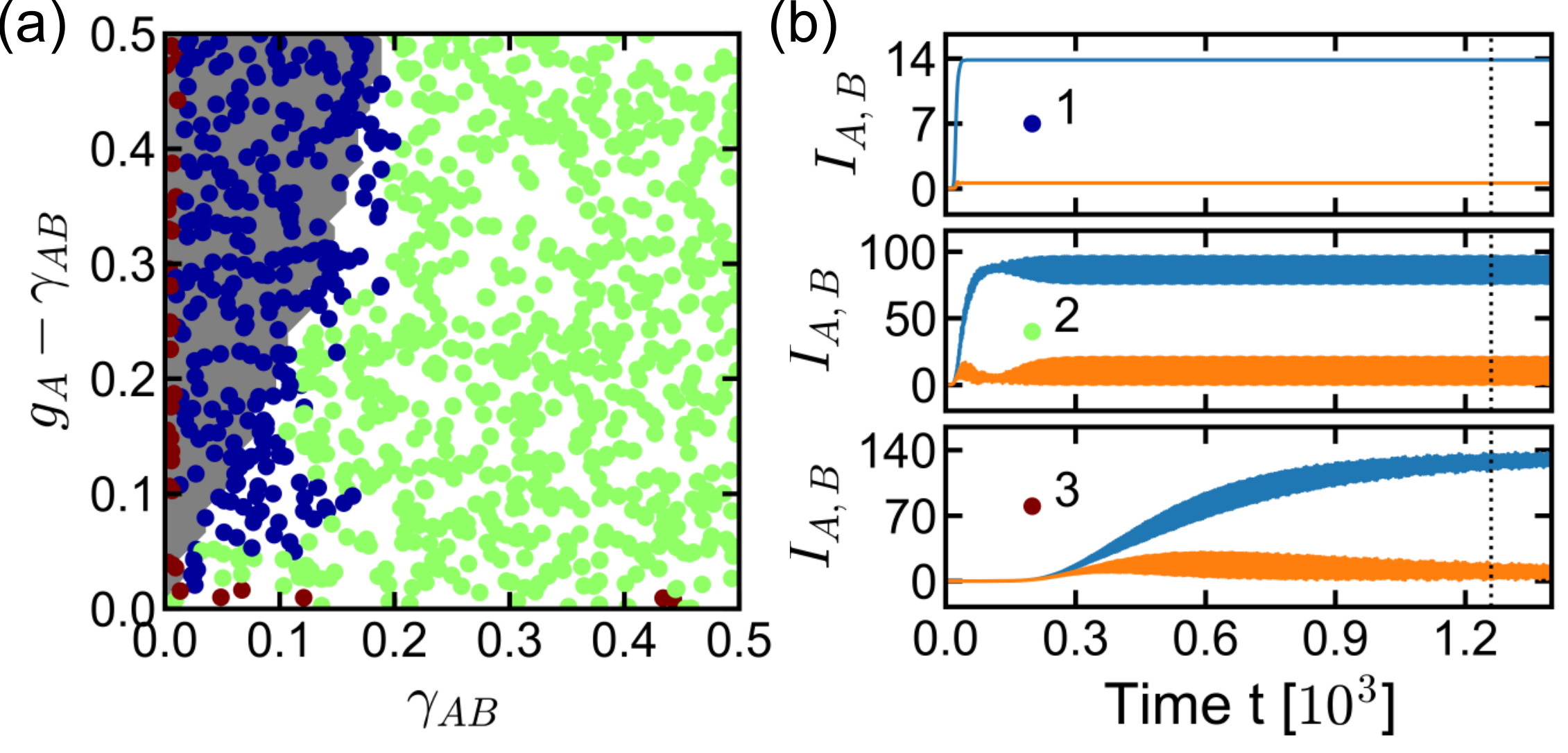}
\caption{
%\textbf{Phase diagram derived based on top-down adaptive representation classification.}
\textbf{Representation classification based on a top-down adaptive library.}
(a) Phase diagram obtained using the top-down classification strategy with an initial library composed of $J=60$ regimes randomly chosen and $\gamma_{th} = 0.75$.
(b) Representative total intensity $I_A$ (and $I_B$) of the A (and B) sublattice in blue (and orange) for the different regimes. 
The black vertical dotted line indicates the starting time from which the bases are generated.
}
\label{fig:phase_diagram_top_down}
\end{figure}

The top-down representation classification strategy is to classify the time series according to a large library of bases, and only then merge the equivalent identified regimes via the calculated alignment subspace $\gamma_{ij}$ and the equivalence relation Eq.~\eqref{eq:top_down}.
Figure~\ref{fig:phase_diagram_top_down} shows the phase diagram obtained from the top-down algorithm that started with a library composed of $J = 60$ regimes randomly chosen, along with the representative of each phase.
We observe in Fig.~\ref{fig:phase_diagram_top_down}(a) that the derived phase diagram is able to distinguish the non-oscillating [top panel of Fig.~\ref{fig:phase_diagram_top_down}(b)] and oscillating regimes [middle panel of Fig.~\ref{fig:phase_diagram_top_down}(b)].
In addition, a third regime corresponding to a transient regime [bottom panel of Fig.~\ref{fig:phase_diagram_top_down}(b)] is found close to the $\gamma_{AB}=0$ or $g_A - \gamma_{AB}=0$ axis.
This transient regime indicates that a longer simulation time might be needed to be considered either in the oscillating or non-oscillating regimes.

However, we can see that the derived phase diagram is still failing in the low $\gamma_{AB}$ and low $g_A - \gamma_{AB}$ region (bottom-left region of the present phase diagram), where some time series are interpreted as non-oscillating instead of oscillating regime.
This shows the limitation of this method where the initially constructed library may lack some of the paths that may connect similar bases.
For example, regimes $i$ and $k$ might not be similar enough to be considered as equivalent directly [Eq.~\eqref{eq:top_down}] but are both equivalent to the regime $k$, \emph{i.e.}, $i \sim j$ and $j \sim k$, which is missing in the library.
The natural workaround would be to increase the initial library size and ensure that the regimes in the library have no missing paths, as we will see in the next section.

\begin{figure}
\center
\includegraphics[width=\columnwidth]{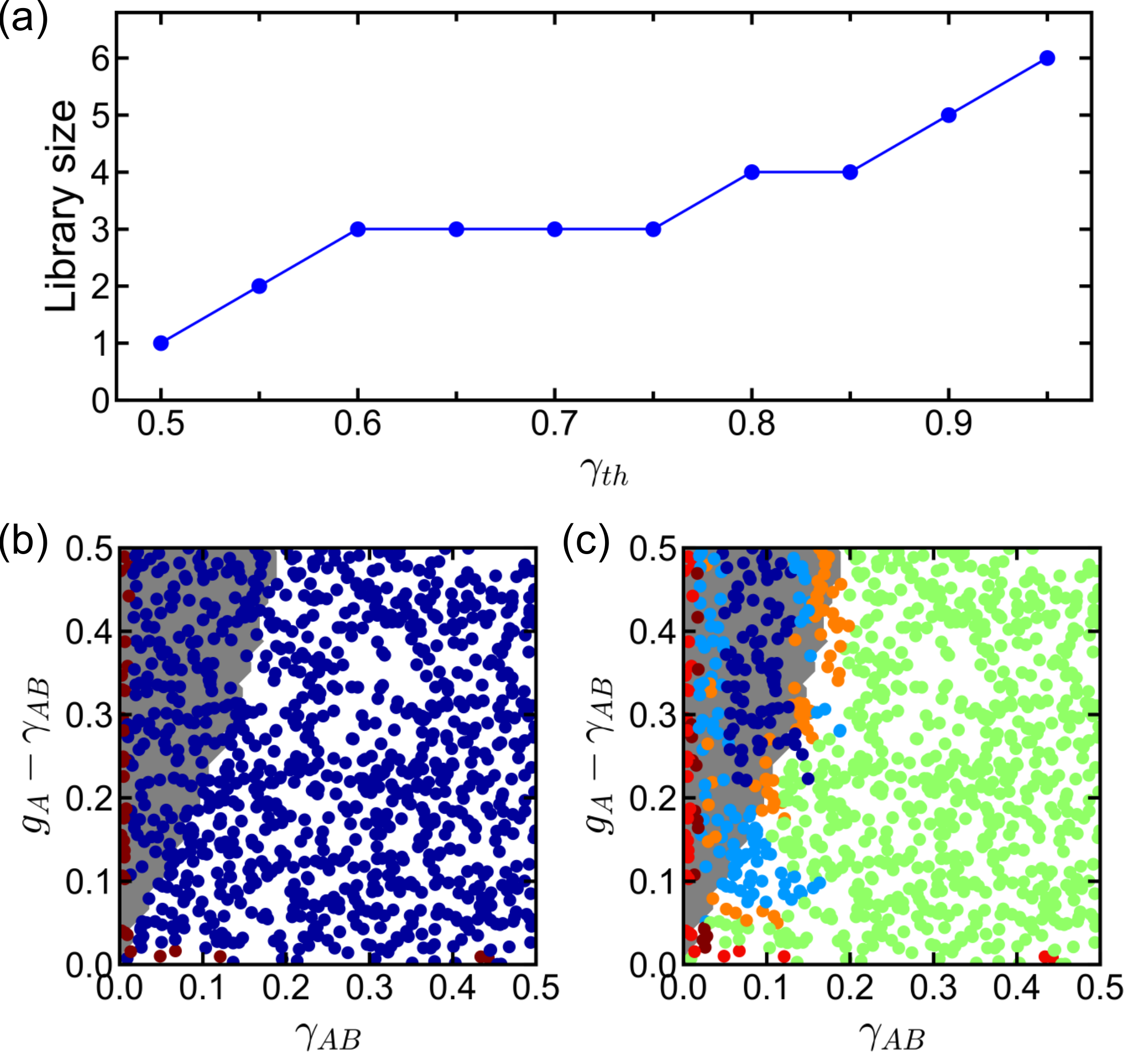}
\caption{
\textbf{$\gamma_{th}$-hyper-parameter dependency.}
(a) Library size against $\gamma_{th}$.
Phase diagrams derived using the top-down classification strategy with (a) $\gamma_{th} = 0.55$ and (b) $\gamma_{th} = 0.95$.
The initial library is composed of $J=60$ regimes randomly chosen.
}
\label{fig:lib_top_down_transition}
\end{figure}

The hyper-parameter $\gamma_{th}$ is an important quantity in the algorithm since it dictates which regimes are equivalent or not.
A low threshold $\gamma_{th}$ will easily merge regimes while a high $\gamma_{th}$ will barely reduce the size of the library as depicted in Fig.~\ref{fig:lib_top_down_transition}(a). 
The threshold is here arbitrarily chosen based on Fig.~\ref{fig:lib_top_down_transition}, and based on the refinement of the desired library.
%
%For example, we can see in Fig.~\ref{fig:lib_top_down_transition}(b) that the derived phase diagram with $\gamma_{th} = 0.65$ has three different phases corresponding to the non-oscillating, oscillating and transient phases.
%
For example, we can see in Fig.~\ref{fig:lib_top_down_transition}(b) that the derived phase diagram with $\gamma_{th} = 0.55$ has two different phases.
For this coarse threshold, the transient regime is identified but the distinct non-oscillating and oscillating phases are merged together into a single phase.
On the other hand, with the same library as in Fig.~\ref{fig:lib_top_down_transition}(b), Figure~\ref{fig:lib_top_down_transition}(c) displays the obtained phase diagram for a finer threshold value $\gamma_{th} = 0.95$.
The plot shows that the algorithm separates the parameter space into several regimes which can be grouped into four main regimes.
In addition to the non-oscillating, the oscillating and the transient regimes, there is a regime corresponding to the transition between the two topological phases.
Besides, this finer description allows us to see distinct sets of modes in the oscillating parameter space region [dark blue and light blue dots in  Fig.~\ref{fig:lib_top_down_transition}(c)].
%
%The analysis of these distinct oscillating regimes found from the data-driven classification is interesting and is left for a future study.
%
Nevertheless, we observe again that the initial library misclassifies some of the non-oscillating time series most likely because of some missing paths, as said previously.
%

%One should note that, compared to standard clustering algorithms which use some metrics to cluster the samples based on some features, the clustering algorithm proposed here simply build, effectively, a connectivity graph from which the number of components of the obtained graph gives us the number of clusters or phases.
%
%However, it would be interesting, as a future work, to build upon the distance between regimes defined as $\gamma_{ij}$ to use in conjunction with standard clustering algorithm such as k-means clustering, spectral clustering, hierarchical clustering, etc~\cite{FabianPedregosa2011}.
%

%=========================
%=========================

%\subsection{Representation classification from bottom-up adaptive library basis}
\subsubsection{Bottom-up adaptive library}

We propose an alternative and dual approach which considers fewer samples in the library.
The core idea of this bottom-up approach is then to add samples on the fly during the classification of the given sample if the library is not good enough.

Here, the library is considered to be good enough if the maximal projection of the measurement onto the regimes' subspace is high enough.
In other words, the library is said to be good enough if the worst relative reconstruction error, $\epsilon$, is low enough:
\begin{equation}
\epsilon < \epsilon_{th}
,
\end{equation}
where $\epsilon_{th}$ is the hyper-parameter which decides the threshold quality of the library and
\begin{equation}
\epsilon := \max_{j=1,\ldots,J} \frac{\norm{P_j y(t) - y(t)}_2}{\norm{y(t)}_2} 
,
\end{equation}
with $\norm{\cdot}_2$ the $L_2$-norm of a vector.
Supplementary section~SIV gives more details on the bottom-up library generation principle.

The advantage of this bottom-up approach is the full exploration of the parameter space region and the automatic construction of a library based on its quality.
This method does not suffer from the randomly chosen samples used to construct the library, and the library composition is not restricted to a narrow parameter space region.
Using a good enough library quality, the algorithm should therefore be able to sort the missing paths issue in the top-down method.

\begin{figure}
\center
\includegraphics[width=\columnwidth]{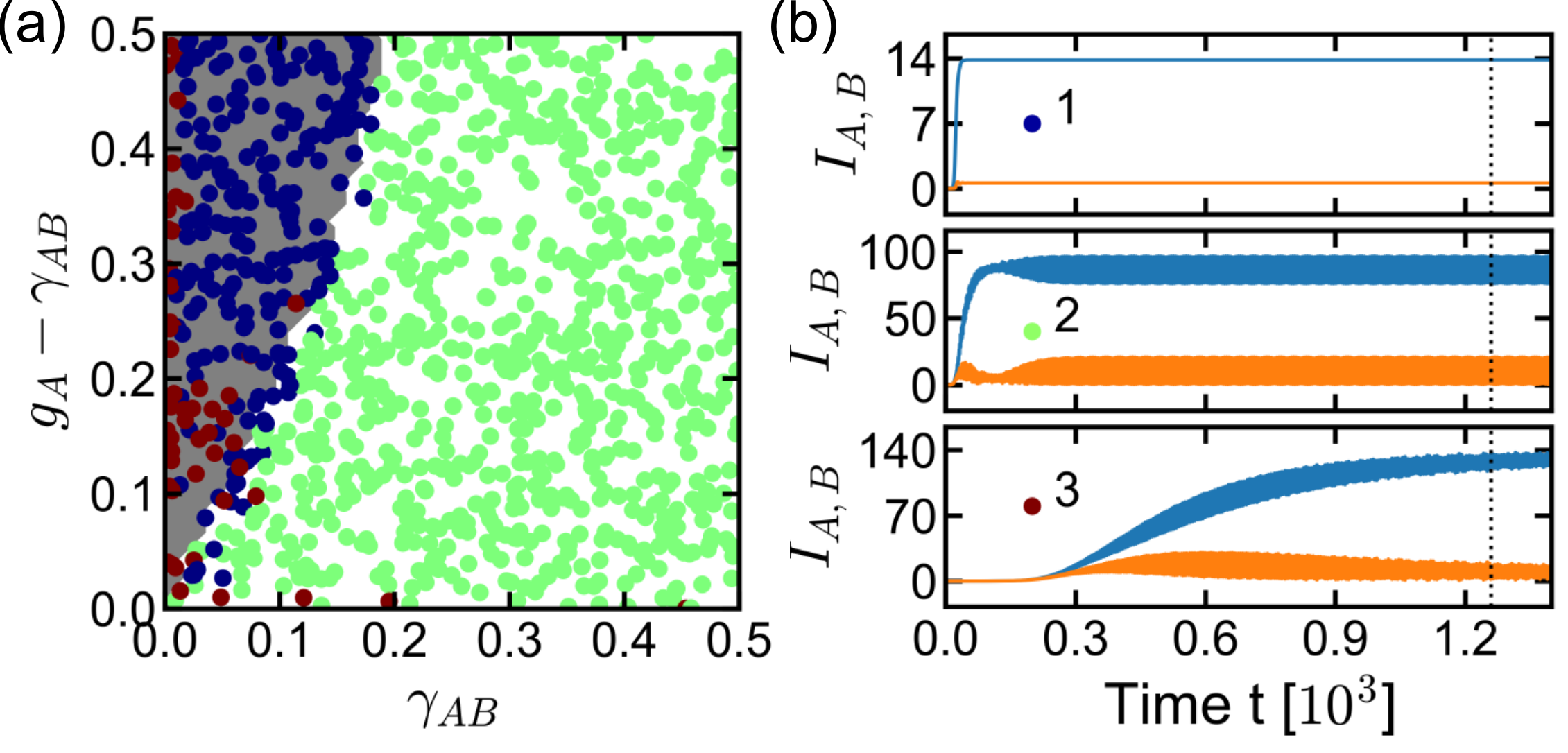}
\caption{
%\textbf{Phase diagram derived based on bottom-up adaptive representation classification.}
\textbf{Representation classification based on a bottom-up adaptive library.}
(a) Phase diagram obtained using the bottom-up classification strategy with a starting library composed of a single regime randomly chosen, $\epsilon_{th} = 0.005$ and $\gamma_{th} = 0.75$.
(b) Representative total intensity $I_A$ (and $I_B$) of the A (and B) sublattice in blue (and orange) for the different regimes.
}
\label{fig:phase_diagram_bottom_up}
\end{figure}

The bottom-up representation classification scheme consists of classifying the time series according to a given library or adding this sample into the library if the library is not good enough, and only then merging the different phases obtained into groups of equivalent regimes using the top-down method.
Figure~\ref{fig:phase_diagram_bottom_up} depicts the phase diagram derived from the bottom-up classification algorithm with a starting library composed of a single regime.
Similarly to the top-down approach in Fig.~\ref{fig:phase_diagram_top_down}(a), we observe three distinct regimes corresponding to the non-oscillating, oscillating and transient regimes [Fig.~\ref{fig:phase_diagram_top_down}(b)].
Nevertheless, the obtained phase diagram now better predicts the regimes.
%
%We do not have misclassifications of non-oscillating and oscillating regimes which were due to missing paths in the library.
%
The misclassifications of the non-oscillating and oscillating regimes, which were due to missing paths in the library, are now reduced, and only very few dots are not correctly identified due to being close to the topological transition boundary.
Likewise, the transient points are indications of longer simulations needed because of the long transient time.

\begin{figure}
\center
\includegraphics[width=\columnwidth]{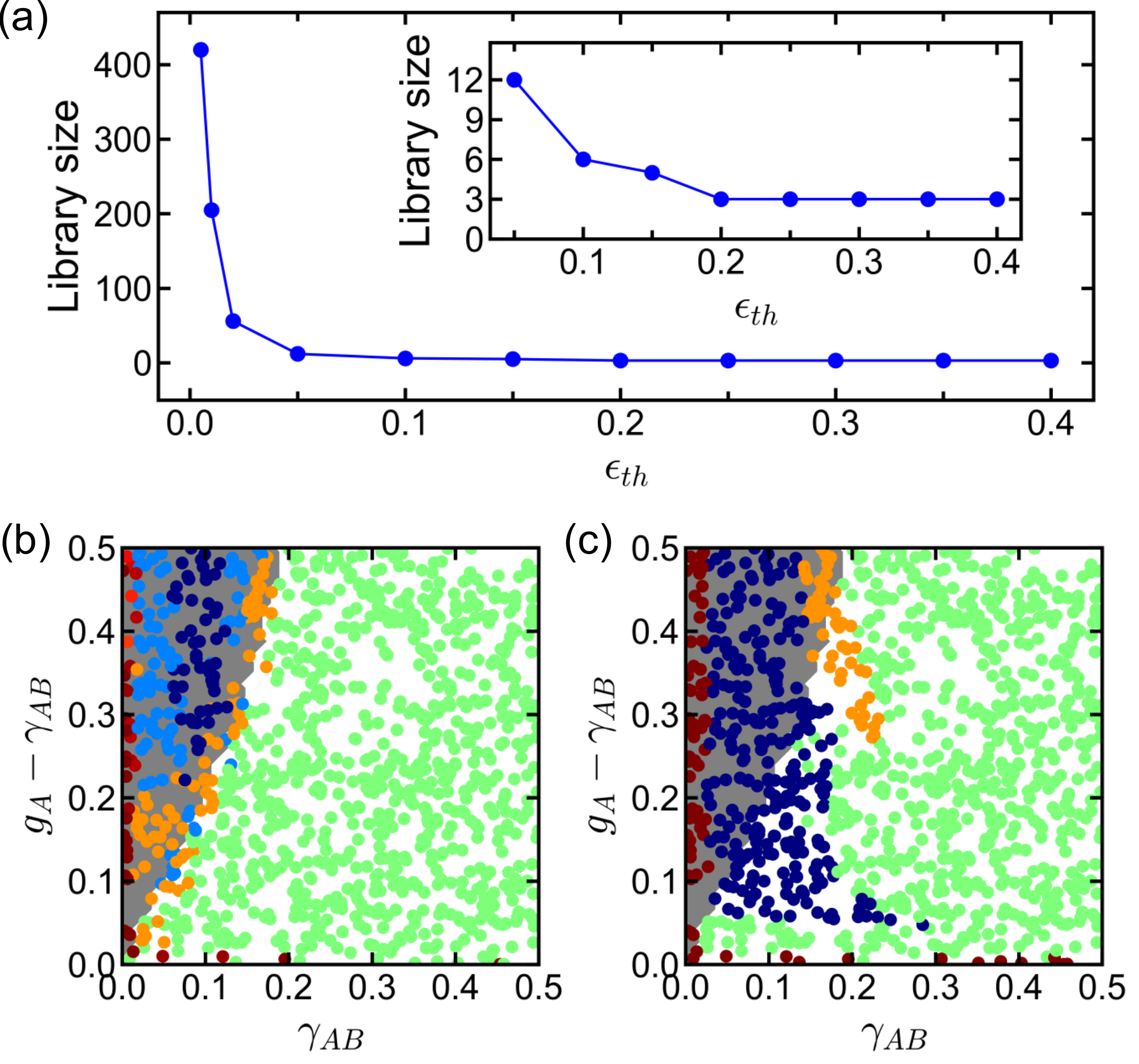}
\caption{
\textbf{$\epsilon_{th}$-hyper-parameter dependency.}
(a) Library size against $\epsilon_{th}$.
The inset is a zoom-in of the plot.
Phase diagram derived using the bottom-up classification strategy with $\gamma_{th} = 0.95$ and (b) $\epsilon_{th} = 0.005$ and (c) $\epsilon_{th} = 0.05$.
The initial library is composed of a single regime randomly chosen.
}
\label{fig:lib_bottom_up_transition}
\end{figure}

Along with the $\gamma_{th}$ hyper-parameter, the threshold hyper-parameter $\epsilon_{th}$ is an important parameter since it tells us whether we want to add or not a given sample into the library.
We observe in Fig.~\ref{fig:lib_bottom_up_transition} that a low threshold $\epsilon_{th}$ will add many samples to the library, whereas a high $\epsilon_{th}$ will not add samples to the library at all.
The threshold value $\epsilon_{th}$ is, again, arbitrarily chosen but with a preference for a high-quality library, \emph{i.e.}, low $\epsilon_{th}$, in order to avoid missing paths.
For example, we can see in Fig.~\ref{fig:lib_bottom_up_transition}(c) the phase diagram derived using $\epsilon_{th} = 0.05$ (and $\gamma_{th} = 0.95$), namely with a library that gives less than $5 \%$ of the reconstruction error of the measurement.
This set of hyper-parameter gives four main regimes that seem to correspond to the non-oscillating, oscillating, transition and transient regimes.
Yet, there is some misidentification of the two topological phases most likely because of missing paths of the obtained library.
On the other hand, with a better library quality, here $\epsilon_{th} = 0.005$ (and $\gamma_{th} = 0.95$), the missing paths are retrieved and the derived phase diagram correctly predicts the topological phases [Fig.~\ref{fig:lib_bottom_up_transition}(b)].
Figure~\ref{fig:lib_bottom_up_transition}(b) shows that the different regimes, obtained previously with a lower library quality, are now better defined.
The non-oscillating and oscillating regimes are well located in their respective parameter space region, and the transition points follow the transition boundary between the two topological phases.
In addition, the bottom-up representation classification is predicting distinct oscillating modes [dark blue and light blue dots in Fig.~\ref{fig:lib_bottom_up_transition}(b)].

The presence of distinct oscillating modes is an example of new insights given by the data-driven classification method.
Indeed, the complex values of the amplitudes of $x(t)$ are, here, taken into account instead of solely the total intensity of each sublattice $A$ and $B$ as in Ref.~\cite{Malzard2018, Malzard2018a}.
This allows for a finer description of the dynamic pattern based on the whole lattice with the relative phase difference of the sites or the absolute value of amplitudes.

%=========================
%=========================
%=========================

\section{Conclusion}

We have proposed a data-driven approach to identify and classify topological phases of dynamical systems. 
By utilizing the representation classification strategy based on the aDMD, we have successfully drawn the phase diagrams of the domain-wall-type SSH lattice with saturable gain. 
To avoid manual labelling in the classification, we have proposed two automatic library construction schemes: top-down and bottom-up approaches
that merge similar phases in a library or adaptively construct a library according to its quality, respectively.
While the bottom-up method is preferred due to the missing paths issue, both methods allow  exploration of parameter space without any expert knowledge of the complex non-linear system.
Our approach is advantageous in doing the reverse engineering to find new phases because only a small desired parameter space region is required for solving the laser rate equation, applying the bottom-up representation classification strategy and building a starting library.
Our method opens the door for finding novel topological lasing modes by providing new insights into the dynamics of coupled lasers in more complicated settings. 
%

%also provides new insights into the dynamics of the topological lasing mode in the parameter space by a careful examination of the derived phase diagrams.
%
%Via reverse engineering, this can be used as a tool to find novel topological lasing modes in more complicated settings.
%
%Given the rate equations of a lasing system, one would only need to integrate the differential equations in the desired parameter space region and then apply the bottom-up representation classification strategy based on the aDMD and with a starting library composed of a single chosen regime.
%
%Scanning over the hyper-parameters $\epsilon_{th}$ and $\gamma_{th}$ might help finding the optimal values needed for the correct classifications.

%Besides, the decomposition method and/or observables used can be tailored for a suitable need with respect to the system's pattern.
%
%The clustering algorithm proposed can also be extended to use standard clustering methods such as k-means clustering, spectral clustering, hierarchical clustering, etc.

%=========================
%=========================

\section*{Data availability}
The data that support the findings of this study are available from the corresponding
author upon reasonable request.

\section*{Code availability}
The code that support the findings of this study are available from the corresponding
author upon reasonable request.

% \bibliographystyle{unsrt}
%\begin{thebibliography}{9}
%\bibliographystyle{IEEEtran}
%\bibliographystyle{apsrev4-1}

%\bibliography{ref}

\section*{acknowledgments}
We acknowledge the support of the Sêr Cymru II Rising Star Fellowship (80762-CU145 (East)) which is part-funded by the European Regional Development Fund (ERDF) via the Welsh Government.

\section*{Author contributions}
S.W., D.R. and S.S.O. conceived the project. 
S.W developed the theoretical model and performed all the numerical simulations.
All authors contributed to the discussions and the preparation of the manuscript. 

\section*{Competing interests}
The authors declare that they have no competing interests.

\section*{Additional information}
Correspondence and requests for materials should be addressed to Sang Soon Oh.

%\include{supp}

%=========================
%=========================
%=========================
%=========================
%=========================
%=========================

\newpage

\renewcommand{\thesection}{S\Roman{section}}
\setcounter{section}{0}

\renewcommand{\thefigure}{S\arabic{figure}}
\setcounter{figure}{0}

\renewcommand{\theequation}{S\arabic{equation}}
\setcounter{equation}{0}

%\appendix 

%=========================
%=========================
%=========================

\section{Basis generation methods}
\label{sect_supp:basis_generation_methods}

This supplementary section review few methods for generating a basis from a time series.
There are many ways to generate the low-order model of a given dynamical behaviour.
An important quantity is the so-called data matrix $X$ built from the data at hand.
The data matrix is a $(N_s \times N_t)$-matrix that collects the $N_t$ data snapshots $x(t_i)$ into columns:
\begin{equation}
X = \left[ x(t_0), x(t_1), \ldots, x(t_{N_t}) \right]
.
\end{equation}
Here, the vector $x(t_i)$ is chosen to be the complex-valued amplitudes of the modes at the A and B sites. 
Other ``observables", such as the absolute values or the total intensity per sublattices, can be used.
This may give different (better or worse) results and is left for a future study.
The bases can then be constructed using dimensional reduction techniques on the data matrix.
Here, we will cover different methods in order to highlight their importance and limitations for the classification scheme used.
%

%=========================
%=========================

\subsection{Proper orthogonal decomposition}

\begin{figure}
\center
\includegraphics[width=\columnwidth]{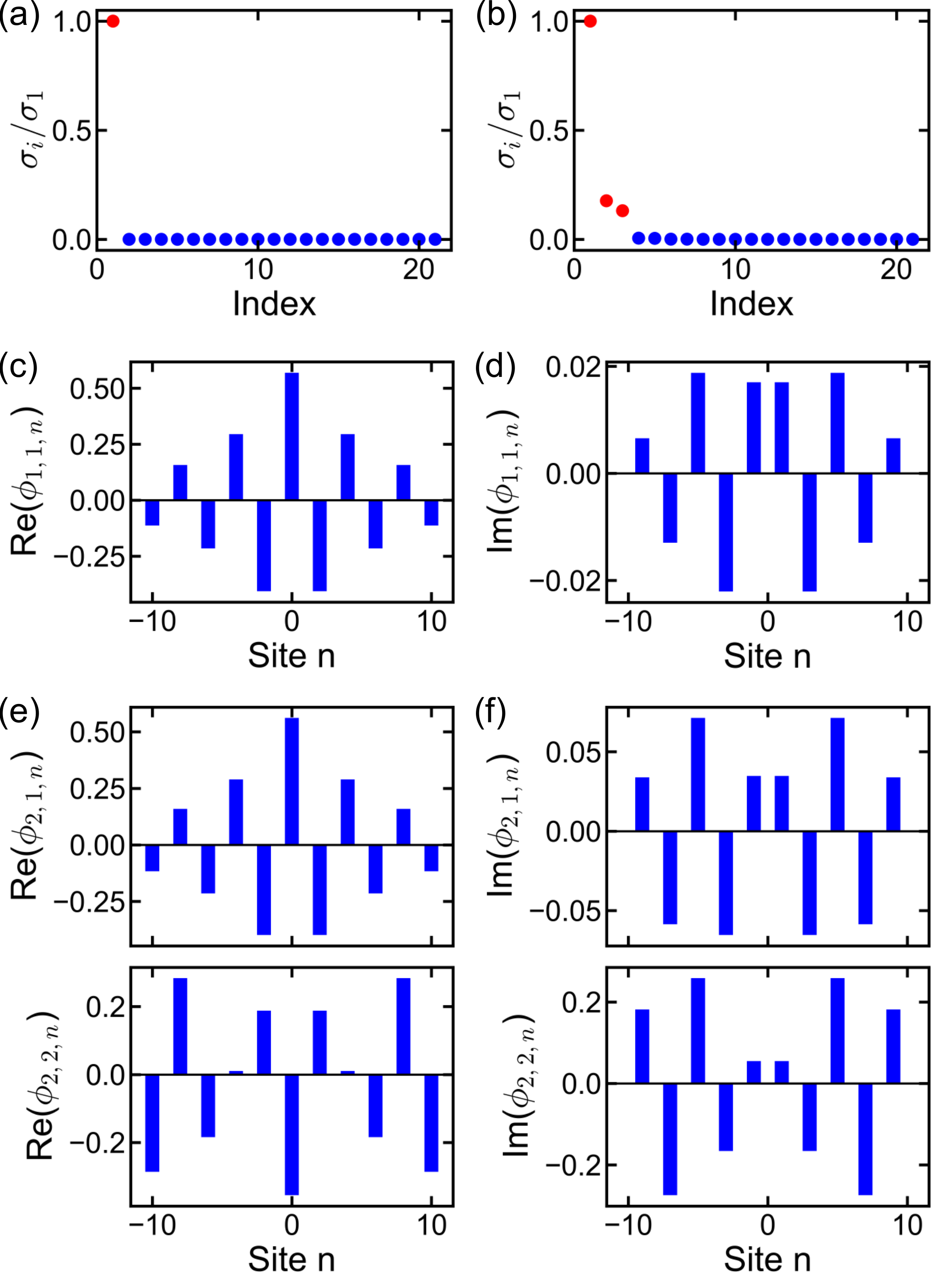}
\caption{
\textbf{POD method.} 
Singular values of the (a) non-oscillating and (b) oscillating regimes.
The red dots correspond to the singular values accumulating $99 \%$ of the total variance of the data.
(c) Real and (d) imaginary parts of the field profile of the first POD mode for the non-oscillating regime.
(e) Real and (f) imaginary parts of the field profile of the (top) first and (bottom) second POD mode for the oscillating regime.
The POD bases have been generated from the time series starting at the $1800$-th time step.
}
\label{fig_supp:basis_pod}
\end{figure}

Proper orthogonal decomposition (POD)~\cite{Proctor2014} is a commonly used tool for dimensional reduction of physical systems. 
This decomposition relies on the singular value decomposition (SVD) of the data matrix, given by:
\begin{equation}
X = U \Sigma V^\dagger
\end{equation}
where $U$ and $V^\dagger$ are $(N_s \times N_s)$ and $(N_s \times N_t)$ unitary matrices, respectively.
$\Sigma$ is a diagonal $(N_0 \times N_0)$-matrix $\text{diag} = (\sigma_1, \ldots, \sigma_{N_0})$, with $N_0 = \min(N_s, N_t)$.
The diagonal entries of $\Sigma$ are the so-called singular values and  are ordered in ascendant order $\sigma_1 > \sigma_2 > \ldots > \sigma_{N_0} \geq 0$.
The SVD gives us two orthonormal bases $U$ and $V^\dagger$ since the matrices $U$ and $V^\dagger$ are unitary matrices.
The columns of $U$ are ordered according to the variance $\sigma_i$ they capture in the data matrix and are called the singular vectors: these are the POD modes that are used in the basis $\Phi$.
Moreover, the POD modes are complex because of the complex data $X$.

For a low-dimensional attractor, the POD basis can be safely truncated at a cut-off value $r$ while retaining the main information of the data matrix.
Explicitly, the SVD reads:
\begin{equation}
\label{eq_supp:svd_explicit}
X_{i m} = \sum_{n=0}^{N_0} U_{i n} \sigma_n V^\dagger_{n m}
,
\end{equation}
and keeping only the $r$ highest terms in the decomposition [Eq.~\ref{eq_supp:svd_explicit}], we have the approximation:
\begin{equation}
X_{i m} \simeq \sum_{n=0}^r U_{i n} \sigma_n V^\dagger_{n m}
.
\end{equation}
This is re-written, in a matrix form, as:
\begin{equation}
X \simeq U_r \Sigma_r V^\dagger_r
\end{equation}
where $U_r$, $\Sigma_r$ and $V^\dagger_r$ are the truncated matrix of $U$, $\Sigma$ and $V^\dagger$, respectively.
Although the cut-off value $r$ can be chosen based on different criteria~\cite{Gavish2014}, $r$ is typically chosen so that the POD modes retain a certain amount of the variance (or energy) $\sigma_X$ in the data, namely:
\begin{equation}
\sum_{i=0}^r \sigma_i > \sigma_X
.
\end{equation}

Figure~\ref{fig_supp:basis_pod} displays the POD method of the non-oscillating and oscillating regimes in the domain-wall SSH lattice with saturable gain [Fig.~\ref{fig:ssh_sat_gain}].
The truncation has been chosen such that $\sigma_X = 99 \%$ of the total variance is retained.
We observe in Figs.~\ref{fig_supp:basis_pod}(a) and \ref{fig_supp:basis_pod}(b) the normalised singular values, and that a single POD mode is retained for the zero-mode-like, whereas three POD modes are needed for the oscillating regime, as marked by the red dots.
The real and imaginary parts of the field profile of the corresponding first few POD modes are plotted in Figs.~\ref{fig_supp:basis_pod}(c) and \ref{fig_supp:basis_pod}(d) and Figs.~\ref{fig_supp:basis_pod}(e) and \ref{fig_supp:basis_pod}(f) for the zero-mode-like non-oscillating and oscillating regimes, respectively.
One can see that the main spatial feature of the zero-mode is captured in the single POD mode obtained after truncation, where the majority of its amplitudes are on the A sublattice.
On the other hand, the POD modes of the oscillating dynamical regime also capture part of the information with some finite amplitudes on the A and B sublattices.

Importantly, in this decomposition [Eq.~\ref{eq_supp:svd_explicit}], SVD is implicitly doing a space-time separation of the data matrix, where the POD modes $U$ contain the spatial information while $V$ have the temporal information at each spatial grid point.
Therefore, the POD modes give a static basis and do not explicitly model the temporal dynamics of the time series. 
This method will therefore most likely fail to identify the correct dynamical regime in the classification step.

%=========================
%=========================

\subsection{Dynamical mode decomposition}

\begin{figure}
\center
\includegraphics[width=\columnwidth]{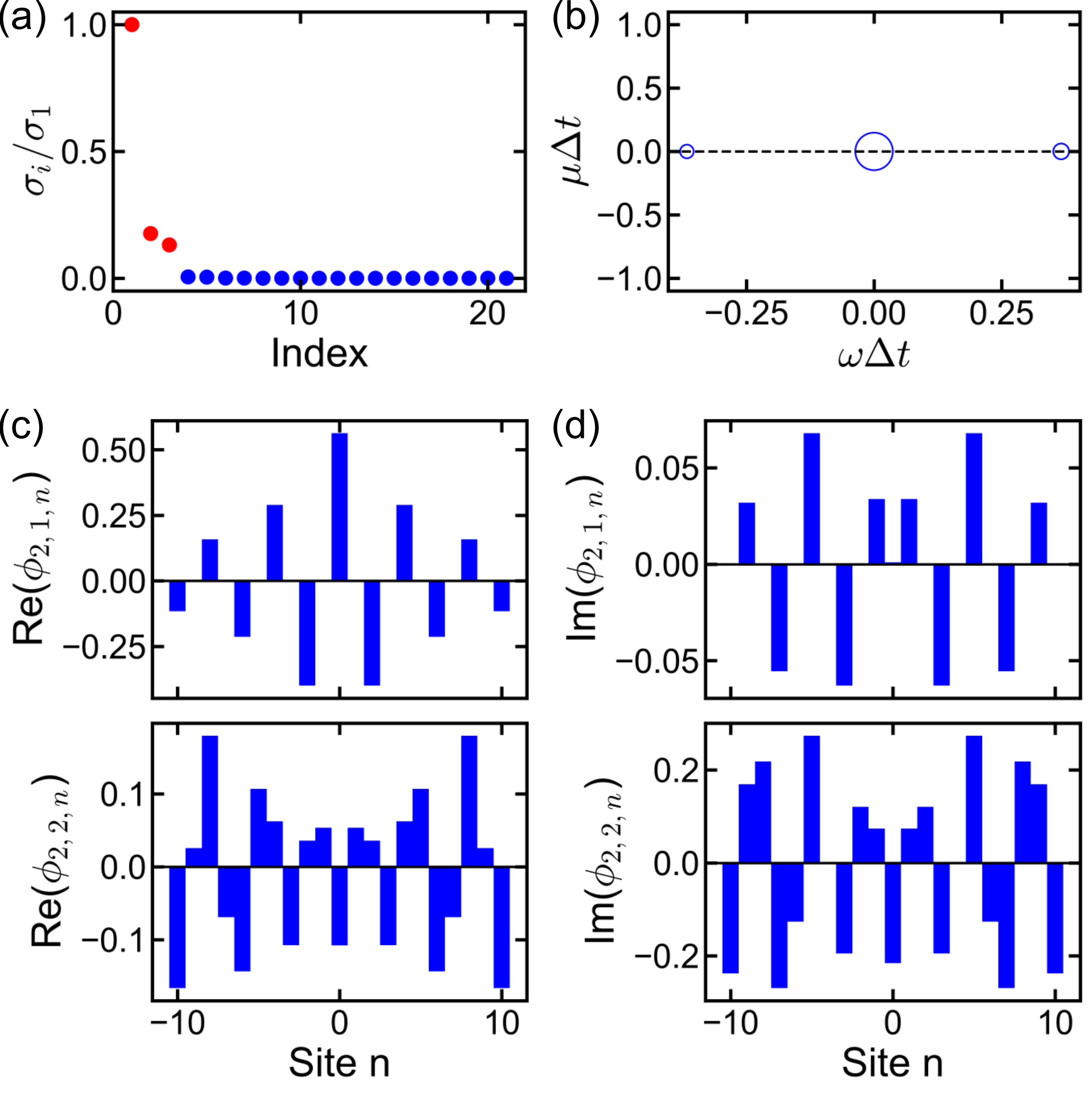}
\caption{
\textbf{DMD method.}
(a) Singular values of the oscillating regime in the DMD.
The red dots correspond to the singular values accumulating $99 \%$ of the total variance of the data.
(b) Plot of the logarithm of the DMD values $\ln(\lambda)$ in the complex plane.
The size of the open circle is proportional to their corresponding singular values.
(c) Real and (d) imaginary parts of the field profile of the (top) first and (bottom) second DMD mode for the oscillating regime.
The DMD basis has been generated from the time series starting at the $1800$-th time step.
}
\label{fig_supp:basis_dmd}
\end{figure}

Dynamical mode decomposition (DMD)~\cite{Schmid2010, Tu2014, Proctor2014} is an alternative to the POD method for learning the dynamics of non-linear systems.
DMD is an explicitly temporal decomposition that takes the sequences of snapshots into account, and is able to derive the spatio-temporal patterns of the data matrix $X$. 
The dynamics of the system are taken into account by considering a linear matrix $A$ that maps a data matrix $X_1$, starting at some time steps $(t_1)$, to the data matrix $X_2$, starting at the next time step ($t_2$).
The matrix $A$ is thus defined as:
\begin{equation}
\label{eq_supp:dmd}
X_2 = A X_1
,
\end{equation}
and the corresponding data matrices are given by:
\begin{equation}
X_1 = \left[ x(t_1), x(t_2), \ldots, x(t_{N_t-1}) \right]
\end{equation}
and
\begin{equation}
X_2 = \left[ x(t_2), x(t_3), \ldots, x(t_{N_t}) \right]
.
\end{equation}
Interestingly, Equation~\ref{eq_supp:dmd} is similar to a linear stability analysis formulation for discrete maps if we think of the stability matrix as the linear matrix $A$.
The DMD method thus consists of solving the following eigenvalues problem:
\begin{equation}
A \Phi = \Phi \Lambda
\end{equation}
where the columns of $\Phi$ are the DMD modes $\phi_i$ and the corresponding DMD eigenvalues $\lambda_i$ are the diagonal entry of $\Lambda$.
The DMD modes $\phi_i$ give us the spatial eigenmodes while their corresponding eigenvalues $\lambda_i$ have their temporal information.
% regarding the growth or decay rate in time or the oscillation frequency.
%
Using a change of units from the data snapshots, observed at every $\Delta t$, to units in time, the eigenvalues are complex-valued scalars:
\begin{equation}
\frac{\ln \left( \lambda_i \right)}{\Delta t} = \mu_i + i \omega_i
\end{equation}
where $\mu_i$ gives the growth (decay) rate if $\mu_i>0$ ($\mu_i<0$) and $\omega_i$ the oscillation frequency of the DMD modes $\phi_i$.

However, the size of the data matrix usually makes the eigendecomposition not feasible.
The goal, here, is therefore to approximate the eigenvalues and eigenvectors of A, using only the data matrices $X_1$ and $X_2$.
The idea is to start by the truncated SVD of $X_1 = U_r \Sigma_r V_r^\dagger$ in which Eq.~\ref{eq_supp:dmd} becomes:
\begin{equation}
X_2 = A U_r \Sigma_r V_r^\dagger
.
\end{equation}
Then the linear matrix $A$ is reduced by considering its projection onto the truncated POD subspace:
\begin{equation}
A_r 
:= U_r^\dagger A U_r 
= U_r^\dagger X_2 V_r \Sigma^{-1}
.
\end{equation}
The eigenvalue problem for $A_r$ is solved with:
\begin{equation}
A_r W = W \Lambda
,
\end{equation}
from which we have the relation:
\begin{equation}
\Phi = X_2 V \Sigma^{-1} W
.
\end{equation}

The key feature of the DMD method is that it decomposes the data into a set of coupled spatio-temporal modes.
The DMD resembles a mixture of the POD in the spatial domain and the discrete Fourier transform (DFT) in the time domain.
%
%The DMD modes are closely related to the POD modes.
%
Figure~\ref{fig_supp:basis_dmd} shows the DMD results for the oscillating regime.
We can, indeed, see in Fig.~\ref{fig_supp:basis_dmd}(a) that the singular values are similar to that of the POD.
Besides, we observe in Figs.~\ref{fig_supp:basis_dmd}(c) and \ref{fig_supp:basis_dmd}(d) that the field profile of the DMD modes closely resembles the POD modes in Fig.~\ref{fig_supp:basis_pod}.
The largest DMD modes not only look similar to the POD modes, but they also contain the oscillation frequencies from $\omega_i$, as in DFT.
The DMD even goes beyond DFT by giving an estimate of the growth (decay) rate in time via $\mu_i>0$ ($\mu_i<0$).
This can be seen by plotting the DMD modes, scaled by their contribution in the decomposition $\sigma_i$, in the frequency plane of $\lambda_i$.
We can see in Fig.~\ref{fig_supp:basis_dmd}(b) that the dynamical regime has a single DMD mode with $\omega_i = 0$ akin to the offset of the oscillation amplitudes, and two DMD modes with opposite $\omega_i \neq 0$ corresponding to the oscillating behaviours. 
All the above three DMD modes have vanishing growth or decay rate $\mu_i=0$.
%

%=========================
%=========================

\subsection{Time-augmented dynamical mode decomposition}

\begin{figure}
\center
\includegraphics[width=\columnwidth]{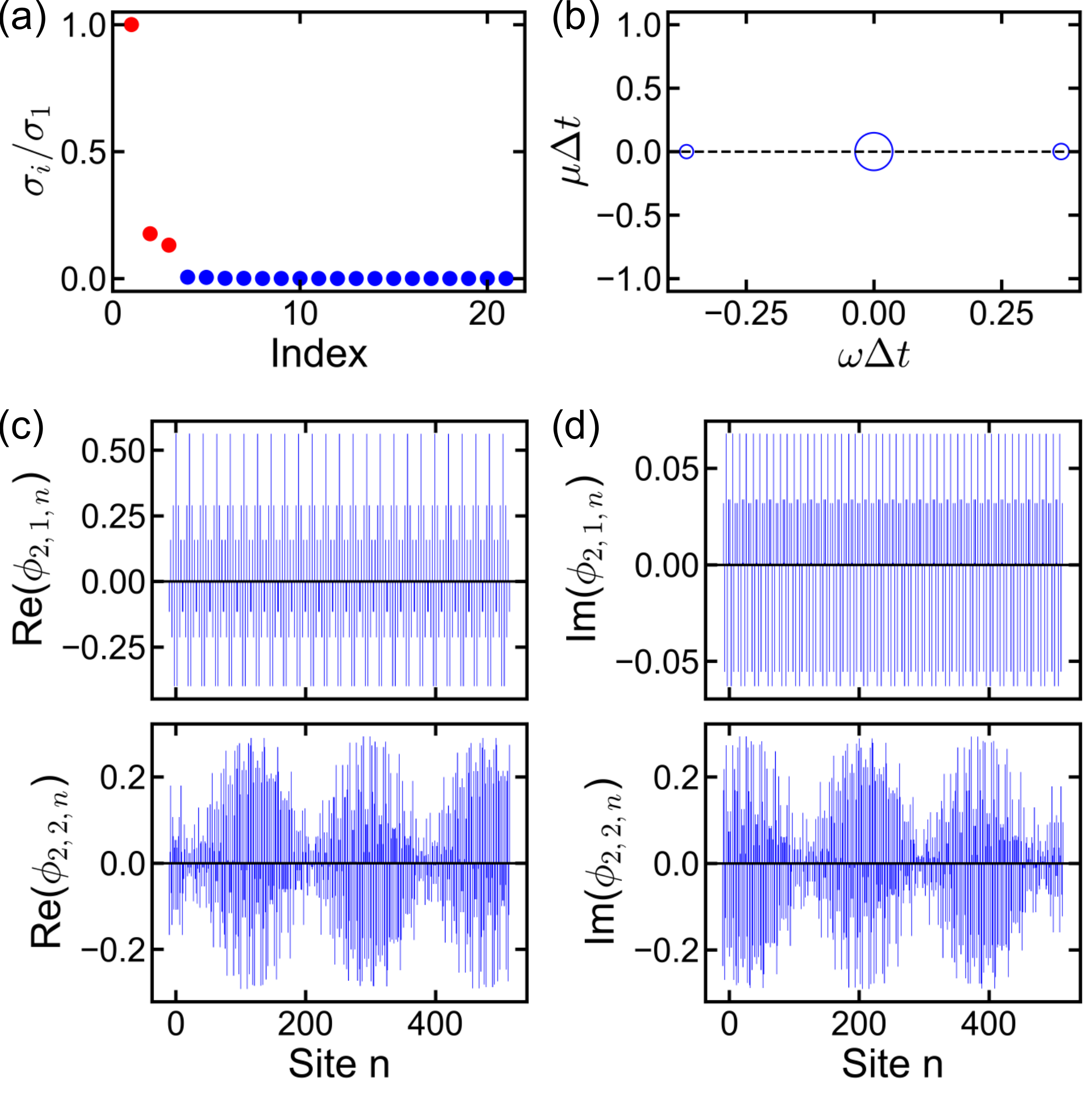}
\caption{
\textbf{aDMD method.}
(a) Singular values of the oscillating regime in the aDMD.
The red dots correspond to the singular values accumulating $99 \%$ of the total variance of the data.
(b) Plot of the logarithm of the aDMD values $\ln(\lambda)$ in the complex plane.
The size of the open circle is proportional to their corresponding singular values.
(c) Real and (d) imaginary parts of the ``field profile" of the (top) first and (bottom) second aDMD mode for the oscillating regime.
Here, by ``Site n", we mean the $n$-th entry of the eigenvector.
The aDMD basis has been generated with $N_w = 25$ from the time series starting at the $1800$-th time step.
}
\label{fig_supp:basis_admd}
\end{figure}

Although the DMD gives the temporal behaviours of the non-linear system, the temporal information is not fully incorporated into the DMD basis $\Phi$ since only the DMD modes are used.
Exploiting the time evolution in the dynamical regime requires the use of DMD modes along with their eigenvalues.
The idea is therefore to incorporate the dynamic information by augmenting the basis $\Phi$~\cite{Kramer2017}.
This time-augmented DMD will be denoted by aDMD in the remaining of this chapter.
Using the defining relation of the eigenvalues $\lambda_i$ as similar to a time evolution operator, \emph{i.e.} multiplying by $\lambda$ is the same as shifting by one time step, we have for a given DMD mode $\phi_i$, its evolution given by $\lambda_i^{N_w} \phi_i$ at $N_w$ time step ahead in time.
Therefore, the time-augmented basis vector reads:
\begin{equation}
\left[
\begin{array}{c}
\phi_i \\ 
\lambda_i \phi_i \\ 
\vdots \\ 
\lambda_i^{N_w} \phi_i
\end{array} 
\right]
.
\end{equation}
%

% The basis decomposition for a single regime [Eq.~\ref{eq_supp:basis_decomposition_single}] is then re-written:
% %
% \begin{equation}
% x(t_i:t_{i + N_w}) = \Phi \beta(t_i:t_{i + N_w})
% \end{equation}
% %
% where the notation $(t_i:t_{i + N_w})$ means we consider a time window from time step $t_i$ to $t_{i + N_w}$.
% %

By considering a time window $N_w$, the time-augmentation of the DMD basis provides us with the dynamical information of the non-linear regime.
Figure~\ref{fig_supp:basis_admd} shows the aDMD results for the same oscillating regime as in Fig.~\ref{fig_supp:basis_dmd}.
We can see that the singular values and aDMD eigenvalues plots [Figs.~\ref{fig_supp:basis_admd}(a) and \ref{fig_supp:basis_admd}(b)] are the same as in the DMD algorithm [Figs.~\ref{fig_supp:basis_dmd}(a) and \ref{fig_supp:basis_dmd}(b)] whereas the ``field profile" of the aDMD modes [Figs.~\ref{fig_supp:basis_admd}(c) and \ref{fig_supp:basis_admd}(d)] carry some temporal evolution information.
In particular, we observe in the top panel of Figs.~\ref{fig_supp:basis_admd}(c) and \ref{fig_supp:basis_admd}(d) the static behaviours of the aDMD mode with $\omega_i=0$.
On the other hand, we can see, in the bottom panel of Figs.~\ref{fig_supp:basis_admd}(c) and \ref{fig_supp:basis_admd}(d), one of the first aDMD modes with $\omega_i \neq 0$ featuring some oscillating behaviour in time.
The size of the basis mode is larger than the plain DMD, and can exhibit its time evolution.
Nevertheless, the graphs do not exactly plot the temporal evolution of the DMD modes since the first $N_s$ entry of the basis state is for the $N_s$ sites; the next $N_s$ for again the $N_s$ sites but at the next time step, etc.
%

%=========================
%=========================

\subsection{Classification results from different decomposition methods}

\begin{figure}
\center
\includegraphics[width=\columnwidth]{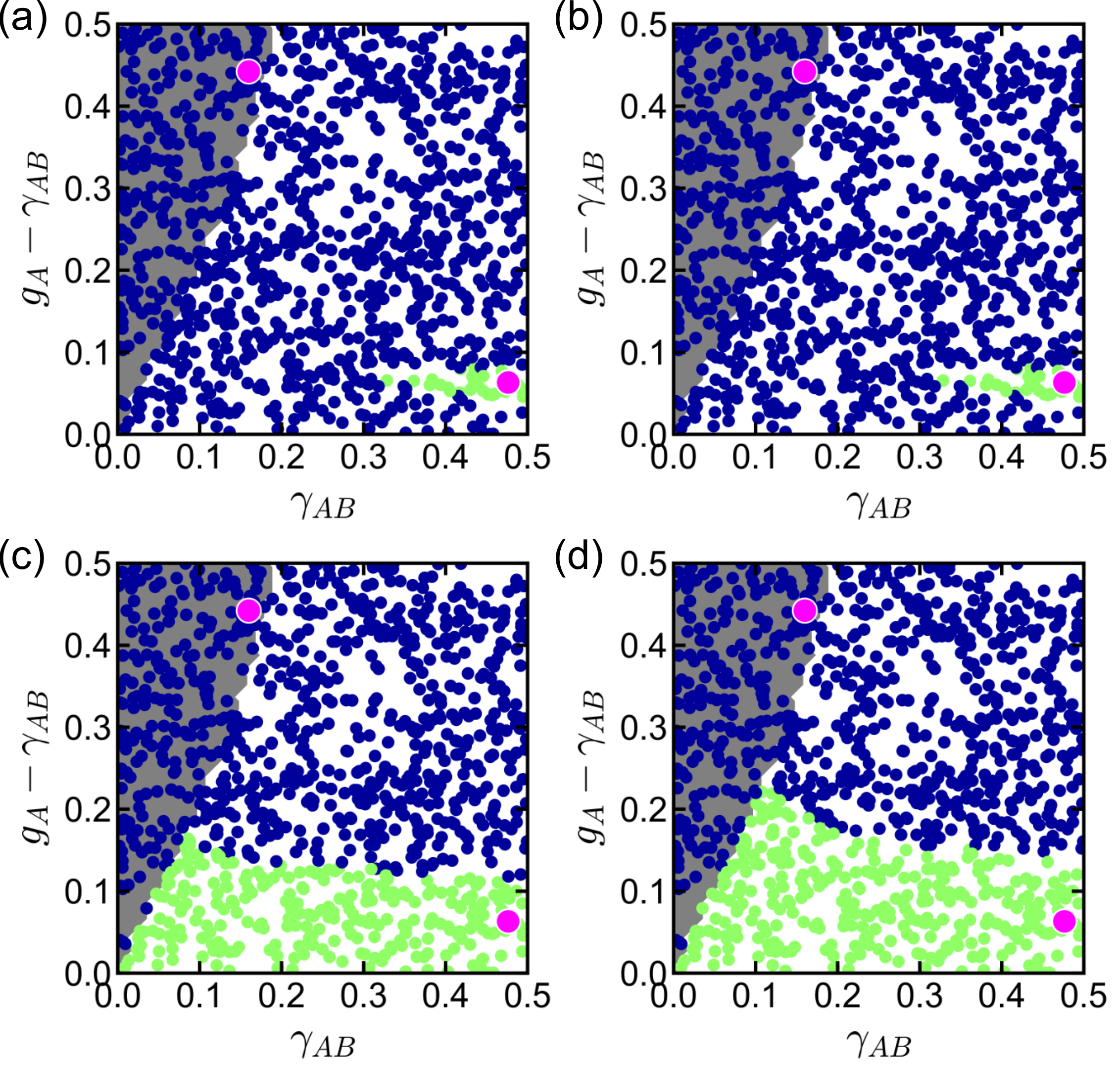}
\caption{
\textbf{Phase diagram derived using different decomposition methods.}
Phase diagrams obtained from a library composed of two regimes (one non-oscillating and one oscillating) from which the bases have been generated using the (a) POD, (b) DMD, (c) aDMD with $N_w=5$ and (d) aDMD with $N_w=25$.
The green and blue dots correspond respectively to the identified non-oscillating and oscillating regimes.
The magenta dots represent the regimes used for the construction of the library.
These are located at ($\gamma_{AB}, g_A-\gamma_{AB}) = (0.48, 0.06)$ and $(0.16, 0.44)$.
The white and grey areas are overlays of the referenced phase diagram obtained in Fig.~\ref{fig:ssh_sat_gain}.
The bases have been generated from the time series starting at the $1800$-th time step.
}
\label{fig_supp:phase_diagram_basis}
\end{figure}

The classification results [Eq.~\eqref{eq:class}] from the decomposition methods reviewed is shown in Fig.~\ref{fig_supp:phase_diagram_basis}.
We observe in Figs.~\ref{fig_supp:phase_diagram_basis}(a) and \ref{fig_supp:phase_diagram_basis}(b) that the phase diagrams fail to correctly predict the distinct dynamical regimes.
This is expected since the POD or DMD modes do not contain enough information about the temporal behaviours.
Besides, the classification for these diagrams is based on a single snapshot ($N_w=0$).
Thus, it is expected the classification fails to capture the correct dynamics since a single snapshot only relies on the spatial pattern of the regime.
On the other hand, we can see in Figs.~\ref{fig_supp:phase_diagram_basis}(c) and \ref{fig_supp:phase_diagram_basis}(d) that the derived phase diagrams have better accuracy when the bases are time-augmented, or equivalently when the classification scheme uses several snapshots.
By increasing the time window in the classification, the derived phase diagram is even better as illustrated in Figs.~\ref{fig_supp:phase_diagram_basis}(c) and \ref{fig_supp:phase_diagram_basis}(d) for $N_w = 5$ and $N_w = 25$, respectively.
The phase diagram will get improved until the time window is large enough to capture the dynamic behaviour.
%

%=========================
%=========================
%=========================

\section{Sparse measurement}
\label{sect_supp:sparse_measurement}

This section detailed the slight change in the methodology in case of sparse sensing.
Sparse sensing is often desirable since the measurement and the data collection can be expensive for a complex system if the space grid is too fine, \emph{i.e.} if $N_s$ is very large.
The compressed measurement $y(t)$ is derived from the full-state measurement $x(t)$ and the measurement matrix $C$:
\begin{equation}
y(t) = C x(t)
,
\end{equation}
where $C$ is a matrix of size ($N_p \times N_s$) with $N_p$ the number of measurements.
Although the measurement matrix $C$ can be represented by some advance and complex mapping~\cite{Candes2006}, here we focus on point-wise measurements, namely the $C_{ij}$ entry in the matrix measurement corresponds to the $i$-th measurement at the $j$-th spatial grid point.
Therefore the compressed basis is given by:
\begin{equation}
\Theta = C \Phi
\end{equation}
where $\Phi$ is the basis obtained from the full-state data collection.
The library of bases for the $J$ distinct dynamical regimes is similarly re-written as:
\begin{equation}
\mathcal{L} = \{ \Theta_1, \ldots, \Theta_J \}
.
\end{equation}
Nevertheless, the size of the current SSH lattice is, here, reasonable and allows us to choose the matrix measurement as the identity matrix $C = 1_{N_s}$.
We will thus use the full-state instead of sparse measurements, but retain the $\Theta$ and $y(t)$ notations to keep the general formalism.
%

%=========================
%=========================
%=========================

\section{Top-down library generation principle}
\label{sect_supp:top_down_lib}

\begin{figure}
\center
\includegraphics[width=\columnwidth]{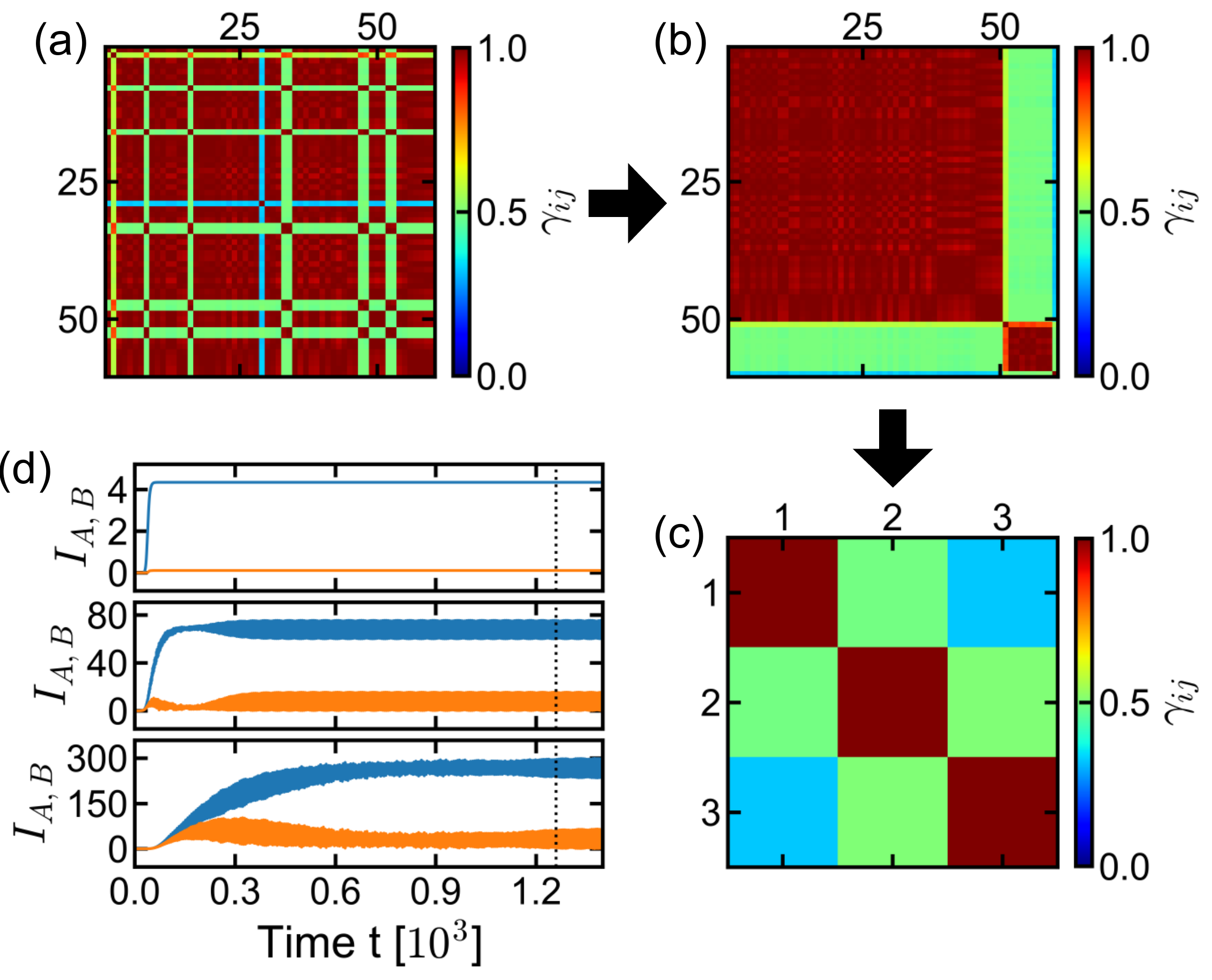}
\caption{
\textbf{Principle of the adaptively top-down library generation.}
(a) Subspace alignment matrix for an initial library composed of $J=60$ regimes.
(b) Subspace alignment matrix from the same library as in (a) but grouped into equivalent regimes.
(c) Subspace alignment matrix from the reduced library after the equivalent regimes are merged with $\gamma_{th} = 0.75$ in (b).
(d) Representative total intensity $I_A$ (and $I_B$) of the A (and B) sublattice in blue (and orange) for the different regimes.
The black vertical dotted line indicates the starting time from which the bases are generated.
The aDMD bases have been generated with $N_w = 25$ from the time series starting at the $1800$-th time step.
}
\label{fig_supp:lib_top_down}
\end{figure}

This section supplement the top-down library generation principle.
The general idea of the top-down construction of the library is to start with a library made of a high number of bases, generated from the time series randomly chosen in the given parameter space region, and then merge them into groups of equivalent regimes.
Figure~\ref{fig_supp:lib_top_down} illustrates the top-down algorithm.
Starting with a library composed of $J$ bases (here $J = 60$), the subspace alignment $\gamma_{ij}$ is computed [Fig.~\ref{fig_supp:lib_top_down}(a)] and then grouped into equivalent regimes according to Eq.~\eqref{eq:top_down} in the main text [Fig.~\ref{fig_supp:lib_top_down}(b)].
Each representative of the regimes is then randomly selected within each group [Fig.~\ref{fig_supp:lib_top_down}(c)].
We plot in Fig.~\ref{fig_supp:lib_top_down}(d) the time series of the representative of each regime.
These regimes are the non-oscillating and oscillating regimes, as well as a third regime which may correspond to a transient regime.
The vertical dashed line in the time series represents the initial time used for constructing the bases in the library.
%

%=========================
%=========================
%=========================

\section{Bottom-up library generation principle}
\label{sect_supp:bottom_up_lib}

\begin{figure}
\center
\includegraphics[width=\columnwidth]{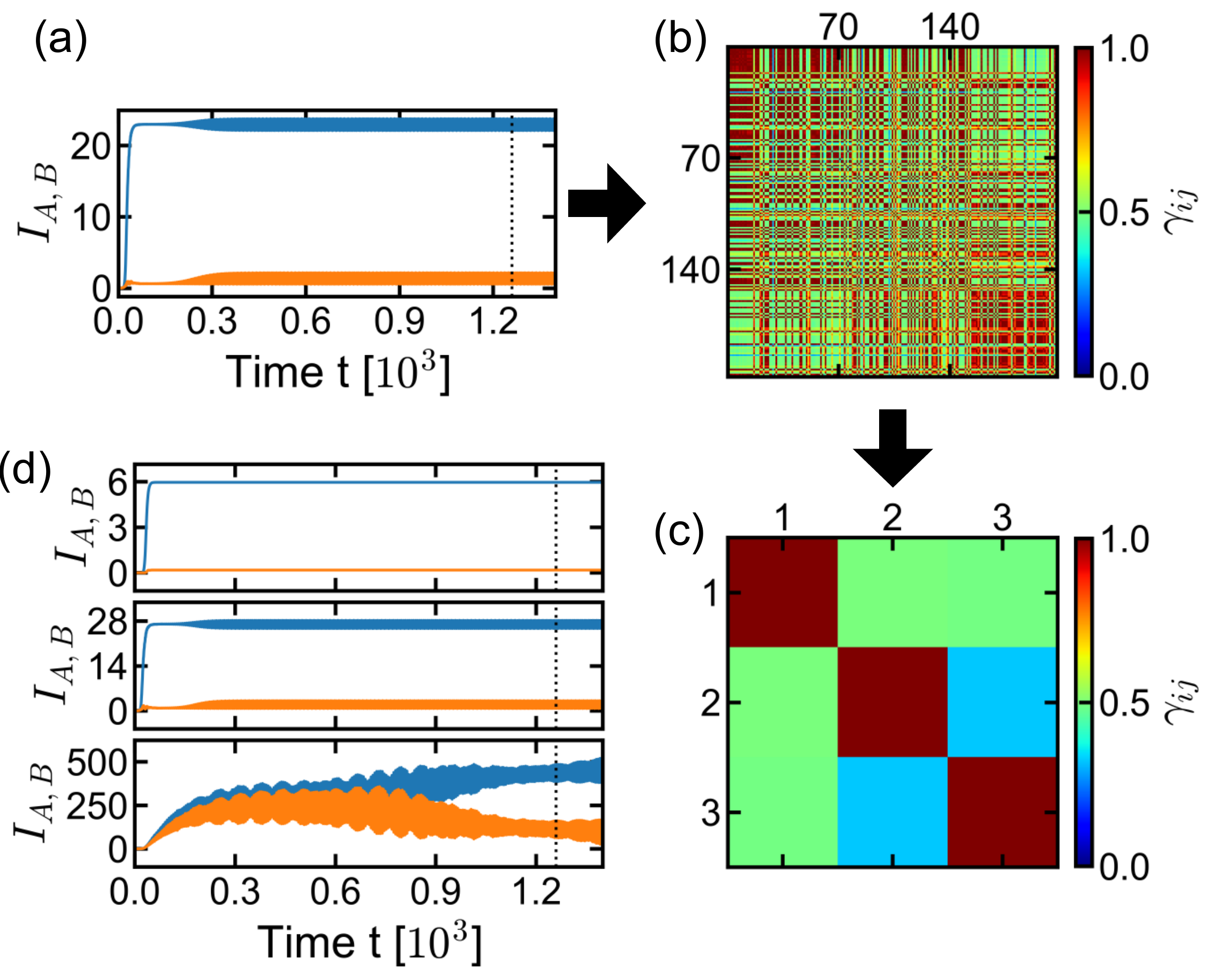}
\caption{
\textbf{Principle of the adaptively bottom-up library generation.}
(a) Total intensity $I_A$ (and $I_B$) on the A (and B) sublattice in blue (and orange) for the single and randomly chosen regime composing the library. 
(b) Subspace alignment matrix from the adaptively constructed library with the single initial regime in (a) and with $\epsilon_{th} = 0.01$.
(c) Subspace alignment matrix from the reduced library after the equivalent regimes are merged with $\gamma_{th} = 0.75$ in (b).
(d) Representative total intensity $I_A$ (and $I_B$) of the A (and B) sublattice in blue (and orange) for the different regimes.
The black vertical dotted line in (a) and (d) indicates the starting time from which the bases are generated.
The aDMD bases have been generated with $N_w = 25$ from the time series starting at the $1800$-th time step.
}
\label{fig_supp:lib_bottom_up}
\end{figure}

This section supplement the bottom-up library generation principle.
Figure~\ref{fig_supp:lib_bottom_up} illustrates the bottom-up methodology proposed.
We start with a single sample in the library, randomly chosen in the parameter space region [Fig.~\ref{fig_supp:lib_bottom_up}(a)].
The library is then adaptively constructed according to the relative reconstruction error $\epsilon$ [Fig.~\ref{fig_supp:lib_bottom_up}(b)].
Finally, with the large library at hand, the top-down approach is used to reduce the library size by merging equivalent regimes [Fig.~\ref{fig_supp:lib_bottom_up}(c)].
The representative of the regimes is plotted in Fig.~\ref{fig_supp:lib_bottom_up}(d) and corresponds to the non-oscillating, oscillating and transient regimes, respectively.

\bibliography{ref}

\end{document}